\documentclass[]{imag-ms-template}

\usepackage{float}
\usepackage{adjustbox}
\usepackage[printonlyused]{acronym}
\usepackage{enumitem}
\usepackage{acronym}
\usepackage{booktabs}
\usepackage{mathtools}
\usepackage{multirow}
\usepackage{makecell}
\usepackage{amssymb}
\usepackage{amsthm}
\usepackage{amsmath}
\usepackage{array}
\usepackage{cleveref}
\usepackage{breakcites} % break citations
\usepackage{xcolor}
\usepackage{upgreek}
\usepackage{longtable}
\usepackage{tablefootnote}

\usepackage{siunitx}
\usepackage{color}

\hypersetup{
    colorlinks,
    linkcolor={black},
    citecolor={black},
    urlcolor={black}
}

\renewcommand{\qed}{\tag*{$\blacksquare$}}

\newcommand{\rev}{\color{black}}
\newcommand{\revtwo}{\color{black}}

\DeclarePairedDelimiter\abs{\lvert}{\rvert}%

\title{V2C-Long: Longitudinal Cortex Reconstruction with Spatiotemporal Correspondence} 

\author{Fabian Bongratz$^{1,2\ast}$, Jan Fecht$^{1}$, Anne-Marie Rickmann$^{1}$, Christian Wachinger$^{1,2}$ 
\\
{\small $^{1}$ Laboratory for AI in Medical Imaging, Technical University of Munich, 81675 Munich, Germany}\\
{\small $^{2}$Munich Center for Machine Learning, Munich, Germany}\\
{\small $^\ast$Correspondence:  fabi.bongratz@tum.de}\\
}

\begin{document} 

\maketitle


\vspace{-5pt}

\begin{abstract}
    Reconstructing the cortex from longitudinal magnetic resonance imaging (MRI) is indispensable for analyzing morphological alterations in the human brain. Despite the recent 
    {\rev advancement}
    of cortical surface reconstruction with deep learning, challenges arising from longitudinal data are still persistent. Especially the lack of strong spatiotemporal point correspondence {\rev between highly convoluted brain surfaces} hinders downstream analyses, as local morphology is not directly comparable if the anatomical location is not matched precisely.
    To address this issue, we
    present \emph{V2C-Long}, the first dedicated deep learning-based cortex reconstruction method for longitudinal MRI.
    V2C-Long exhibits strong \emph{inherent spatiotemporal correspondence} across subjects and visits{\rev, thereby {\revtwo reducing} the need for surface-based post-processing}. We establish this correspondence directly during the reconstruction via the composition of two deep template-deformation networks and innovative aggregation of within-subject templates in mesh space. 
    {\rev {\revtwo We validate} V2C-Long on two large neuroimaging studies, focusing on surface accuracy, consistency, generalization, test-retest reliability, and sensitivity.}
    The results reveal a substantial improvement in longitudinal consistency and accuracy compared to existing methods. In addition, we demonstrate stronger evidence for longitudinal cortical atrophy in Alzheimer's disease than longitudinal FreeSurfer.
\end{abstract}

\begin{figure}[htb]
    \centering
    \includegraphics[width=0.75\textwidth]{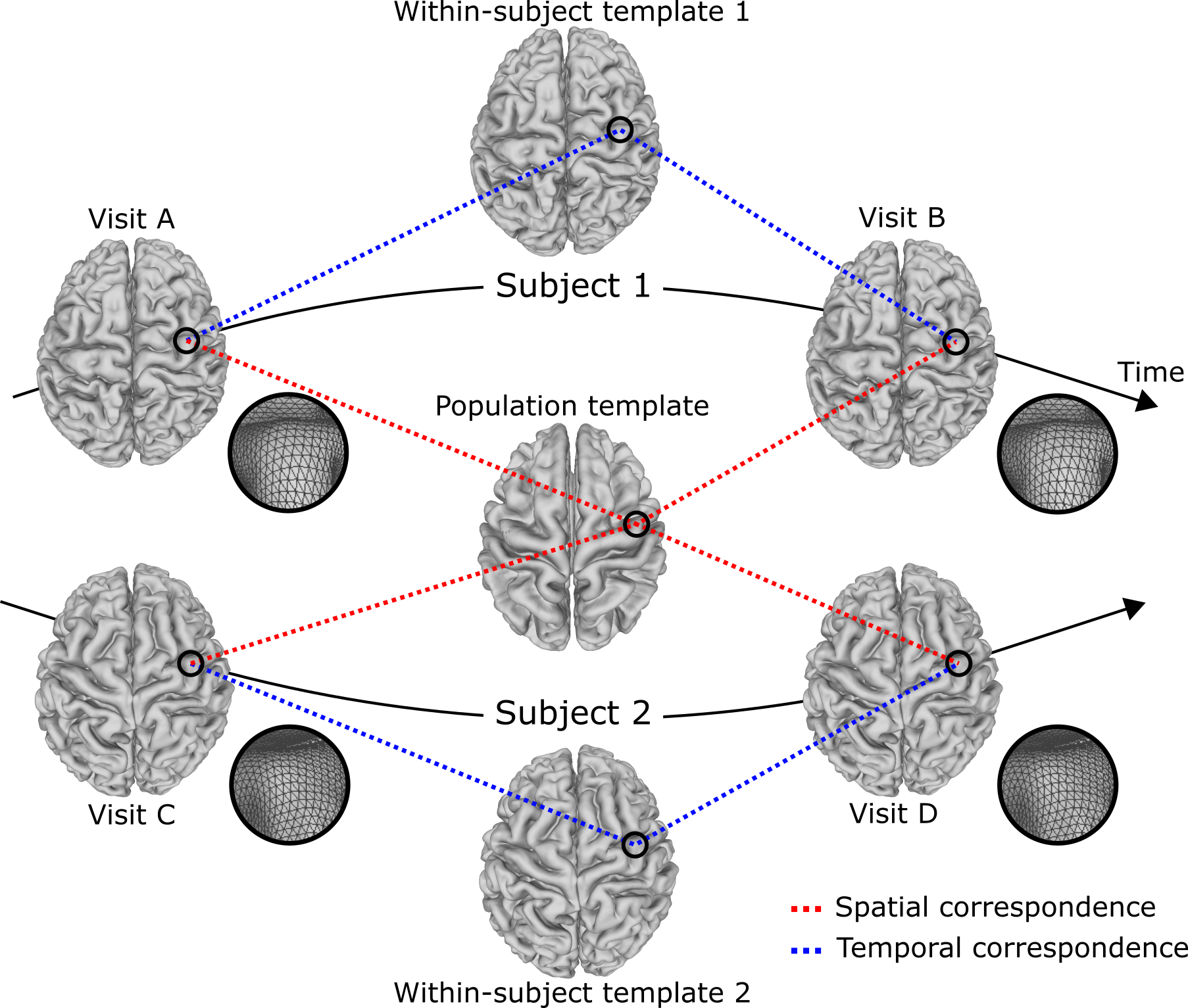}
    \caption{\rev V2C-Long inherently establishes spatiotemporal correspondence of cortical surfaces during the reconstruction. This renders all reconstructed surfaces, i.e., within a certain subject and across subjects, directly comparable on the vertex level.}
    \label{fig:concept}
\end{figure}

\section{Introduction}

The structure of the 
{\rev cortical gray matter}, 
i.e., the thin and tightly folded sheet of neural tissue confined by inner white matter (WM) and outer pial surfaces, can be observed from structural magnetic resonance imaging (MRI). It has direct implications for brain functionality~\citep{Pang2023geometricconstraints}, and local structural measurements like cortical thickness or curvature are important biomarkers to understand brain development~\citep{mills2014longitudinal} and to monitor the progression of brain disorders~\citep{schwarz2016alzheimers,Bachmann2023longchanges,Risacher2010}. 
For accurate \emph{in vivo} measurements, 
triangular meshes representing the WM and pial surfaces are typically extracted from the {\rev MR image} since they are less susceptible to partial-volume effects than voxel-based segmentation~\citep{daleCorticalSurfaceBasedAnalysis}. 
{\rev 
The surface-based representation enhances the sensitivity to subtle changes in cortical morphology, which is crucial for detecting early signs of neurodegenerative diseases and monitoring disease progression. Moreover, the high spatial resolution of triangular meshes allows for high-resolution mapping of cortical features, enabling researchers and clinicians to pinpoint specific regions of interest with great precision. 
}
Recently, the reconstruction of cortical surfaces {\revtwo has been accelerated} from hours to seconds with deep-learning models that can be run on the latest generation of graphics processing units (GPUs)~\citep{cruzDeepCSR3DDeep2020,lebratCorticalFlowDiffeomorphicMesh,gopinathSegReconLearningJoint2021,hoopesTopoFitRapidReconstruction,maCortexODELearningCortical2022}.

These neural networks focus on cross-sectional data, but studying within-subject changes in aging, disease, and treatment relies on longitudinal data.
Processing longitudinal neuroimaging data requires dedicated tools to reduce bias and increase the sensitivity to subtle differences in follow-up visits, typically far below the {\rev image resolution} of 1mm~\citep{reuterWithinsubjectTemplateEstimation2012}. 
To this end, unbiased within-subject templates and homologous points, i.e., corresponding anatomical locations, must be estimated accurately to make a local comparison of brain morphometry possible~\citep{reuterAvoidingAsymmetryinducedBias2011}. More precisely, within-subject templates condense multiple longitudinal snapshots of an individual's brain anatomy into a single geometric model for better within-subject consistency and higher sensitivity to slight morphological alterations. This is especially important for adult brains, where changes are more subtle than in early developmental stages~\citep{bethlehem2022braincharts}.
Within-subject templates further serve as a common ground for visualizations and spatiotemporal statistics. 
{\rev
The established approach for longitudinal cortical surface reconstruction,
}
e.g., as implemented in FreeSurfer~\citep{fischlFreeSurfer2012} and illustrated in Supplementary Figure~2, is to perform a group-wise registration of all of a subject's visits, extract cortical surfaces from the obtained template image, and use these within-subject templates to initialize the reconstruction at each visit. 
However, this approach suffers from several flaws. 
First, although {\rev correspondences between the subject's visits are established, the within-subject templates are not directly comparable across subjects.} This complicates downstream applications, e.g., longitudinal group comparisons. In addition, adding a new scan requires re-running the entire pipeline, starting from the group-wise registration, which is inefficient and, therefore, time- and resource-consuming. Finally, the group-wise registration is prone to bias introduced by variations in voxel intensity or registration asymmetry~\citep{reuterAvoidingAsymmetryinducedBias2011}, challenges that may be overcome with advanced neural deformation networks.

In this work, we introduce V2C-Long --- a novel {\revtwo cortical reconstruction framework} dedicated to longitudinal MRI. V2C-Long establishes strong spatiotemporal correspondence of surface points, i.e., within subjects \emph{and} across subjects, inherently during the reconstruction as illustrated in \Cref{fig:concept}. {\rev The inherent correspondence {\revtwo eliminates the need for spherical inflation} to create correspondences within and across subjects.} {\revtwo To achieve this}, we propose a two-stage approach that builds upon recent advances in template-based cortical surface reconstruction. First, we train a V2C-Flow model~\citep{Bongratz2024v2cflow} to deform {\rev the FsAverage population template~\citep{Fischl1999fsaverage}} to cortical surfaces and leverage the inherently established spatial correspondence to compute within-subject templates directly on the surface level. We then incorporate the within-subject templates into a second deep template-deformation network, thereby establishing strong temporal correspondence among multiple cortical surfaces from an individual. The learning-based approach, together with the parallel execution on the latest generation of GPUs, allows us to compute application-ready longitudinal sequences of cortical surfaces within seconds. Moreover, as the input and output data formats are compatible with existing neuroimaging tools, V2C-Long can be readily integrated into existing image-processing pipelines. {\rev {\revtwo We evaluate} V2C-Long and derived cortical thickness measurements thoroughly using two large neuroimaging studies, ADNI and OASIS, and a test-retest database. We} show that V2C-Long compares favorably with existing methods regarding surface accuracy, longitudinal consistency, and test-retest reliability. In addition, {\revtwo we perform} a downstream analysis of longitudinal cortical thickness in Alzheimer's disease,
{\rev observing more significant evidence}
of group differences in V2C-Long compared to the best alternative approaches.

\section{Related Work}
Originally, the task of cortical surface reconstruction was tackled by a sophisticated combination of segmentation and mesh-extraction algorithms~\citep{daleCorticalSurfaceBasedAnalysis,MacDonald2000,Mangin1995}, with runtimes in the order of hours for a single scan. Recently, deep learning-based cortex reconstruction methods have emerged, allowing for fast surface extraction within seconds on the latest GPUs. These methods can be categorized into {\rev implicit, segmentation-based, and template-based} approaches, depending on the input and output of the utilized neural network. Implicit~\citep{cruzDeepCSR3DDeep2020,gopinathSegReconLearningJoint2021} and segmentation-based approaches~\citep{maCortexODELearningCortical2022,henschelFastSurferFastAccurate2020} first compute an implicit representation of the surfaces, i.e., a signed distance function or occupancy grid, respectively, which is then converted to an explicit mesh with marching cubes~\citep{Lorensen1987marching} or similar algorithms. However, applying marching cubes precludes a direct comparison of the surface meshes on a per-vertex basis and the resulting surfaces come without correspondence.
{\rev
In general, cortical surfaces can still be registered post hoc, but this requires intricate surface inflation and mapping to an icosphere, inevitably introducing area and shape distortions~\citep{yeoSphericalDemonsFast2010,robinsonMultimodalSurfaceMatching2018,sulimanDeepDiscreteLearningFramework2022}.
}

{\rev Although} implicit deep learning-based methods offer a considerable speed-up compared to traditional neuroimaging processing, explicit template-deformation methods~\citep{lebratCorticalFlowDiffeomorphicMesh,hoopesTopoFitRapidReconstruction,cruzCorticalFlowBoostingCortical2022,rickmannVertexCorrespondenceCortical2023,Bongratz2024v2cflow,wickramasingheVoxel2Mesh3DMesh2020,Ma2023cotan} are usually fastest. These methods do not require a conversion of surface representations. Instead, they follow the idea of deformable contours~\citep{Kass1988-ym,Mangin1995} in that they take a generic, topologically correct shape template as input and deform it to individual three-dimensional shape contours based on latent features extracted from the input image. Apart from the fast processing, this approach {\revtwo bears the potential to establish correspondences between the template and the predicted surfaces} through the deformation, enabling direct cross-sectional analyses and brain atlas propagation~\citep{Bongratz2024v2cflow,rickmannVertexCorrespondenceCortical2023}.
{\rev
Different from spherical surface registration, the corresponding points lie directly on the folded cortical surfaces and not on an icosphere.
In V2C-Flow~\citep{Bongratz2024v2cflow} and CorticalFlow~\citep{lebratCorticalFlowDiffeomorphicMesh}/CorticalFlow$^{++}$~\citep{cruzCorticalFlowBoostingCortical2022}, these correspondences are learned in an unsupervised manner with the Chamfer loss and regularization terms. The Chamfer loss does not require ground-truth correspondences during training. V2CC~\citep{rickmannVertexCorrespondenceCortical2023}, on the other hand, takes a supervised approach and computes an L1 loss between the predicted vertices and registered and re-sampled reference surfaces. Lastly, the loss from TopoFit~\citep{hoopesTopoFitRapidReconstruction} can be considered to lie in between the Chamfer and L1 loss. More precisely, TopoFit restricts the Chamfer loss to a certain neighborhood based on the adjacency of vertices in the FsAverage template; again, this approach requires registered and re-sampled reference surfaces.
}

For longitudinal cortex analysis, dedicated computational pipelines~\citep{reuterWithinsubjectTemplateEstimation2012,Li2012consistentreconstruction,Ashburner2013longMRI} were developed and implemented into widely used neuroimaging tools like FreeSurfer~\citep{fischlFreeSurfer2012} and CAT12~\citep{Gaser2022cat}. The common approach is to estimate a subject-specific mean or median image via group-wise registration as an intermediate step to obtain within-subject template surfaces. These within-subject templates are then warped to the contours of the individual scans. Thereby, the vertices within a longitudinal sequence are matched, whereas the within-subject templates from different individuals are not comparable on the vertex level. Hence, error-prone and time-consuming surface inflation, registration, and re-sampling is inevitable for longitudinal group analyses in these approaches~\citep{Fischl1999inflation}. 

More broadly, {\rev deep-learning methods also exist that learn abstract latent representations} from longitudinal images~\citep{kim2023learning,ren2022localspatiotemporal}. These methods, however, are not designed to work with non-Euclidean representations such as brain surfaces. A notable exception is the recent work by \citet{Zhao2024longitudinallyconsistent}, which addresses longitudinal cortical parcellation but, unfortunately, is limited to a canonical spherical representation of cortical surfaces and, hence, relies on an accurate reconstruction in the first place.

\section{Materials and Methods}

\begin{figure}[t]
\centering
\includegraphics[width=1.0\textwidth]{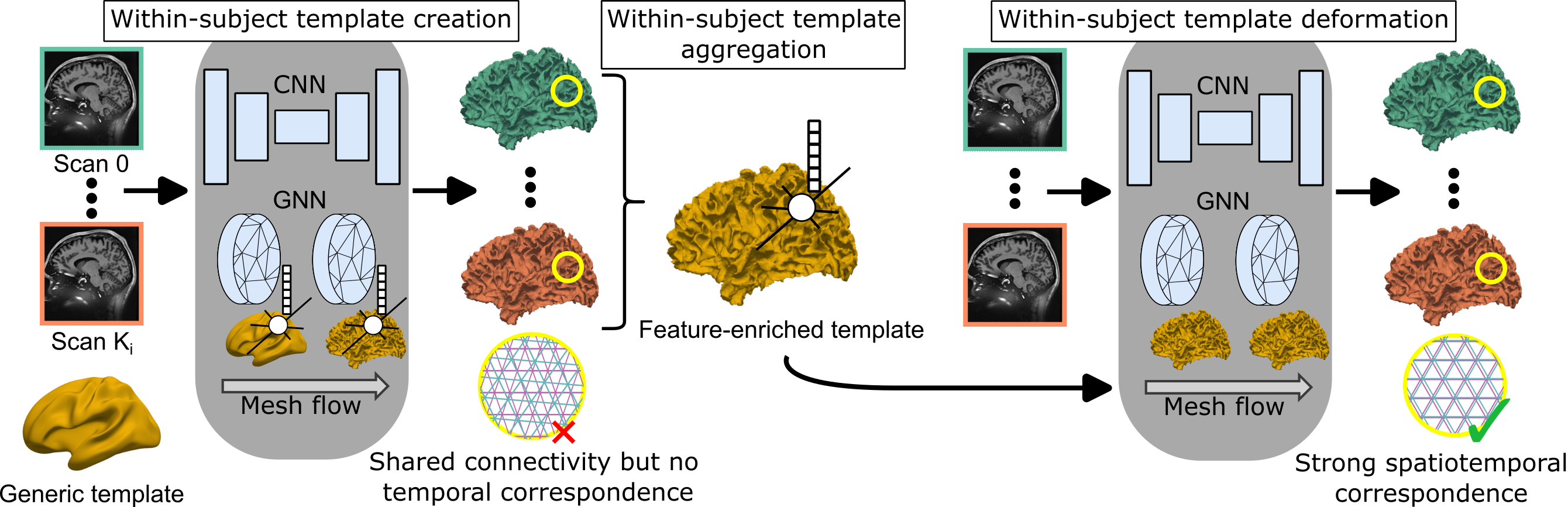}
\caption{\rev Architecture of V2C-Long. From a sequence of 3D brain MRI scans, V2C-Long computes a within-subject template, enriches it with vertex features from the template-creation model, and deforms it to cortical surfaces with strong spatiotemporal point correspondence.}
\label{fig:v2clong}
\end{figure}

\Cref{fig:v2clong} depicts the architecture of V2C-Long. As input, V2C-Long takes a sequence of MR images from a certain subject. As output, WM and pial surfaces with spatial, i.e., across subjects, and temporal, i.e., within a subject, correspondence of anatomical surface points are computed for each input image, cf.~\Cref{fig:concept}. We describe the architecture of the neural network, the within-subject template creation, and the within-subject template deformation in \Cref{sec:architecture,sec:creation,sec:deformation}, respectively.

\subsection{Data and preprocessing}
We obtained data for the preparation of this article from the Alzheimer's Disease Neuroimaging Initiative (ADNI) database (\url{http://adni.loni.usc.edu}), containing subjects with Alzheimer's disease, mild cognitive impairment, and cognitively normal individuals. {\rev We used longitudinal T1w MRI scans (1.5T and 3T)}, split the images on the subject level, and stratified our splits according to sex, age, {\rev initial diagnosis}, and number of visits per subject ($\text{mean}=4.4$ visits, $\text{SD}=1.9$). This resulted in 3,745, 594, and 1,094 scans for training, validation, and testing, respectively. Further, we used 288 scans (100 subjects) from the longitudinal OASIS-3~\citep{lamontagneOASIS3LongitudinalNeuroimaging2019} study to evaluate the generalization of the trained models to an external data set. As a reference standard, we used surfaces reconstructed by FreeSurfer v7.2~\citep{fischlFreeSurfer2012}, cross-sectional processing, which has become a standard in the field~\citep{cruzDeepCSR3DDeep2020,maCortexODELearningCortical2022,lebratCorticalFlowDiffeomorphicMesh,hoopesTopoFitRapidReconstruction}. As model input, we provided the \texttt{orig.mgz} files, mapped to \texttt{MNI152} standard space (1mm, $182\times218\times182$ voxels) by affine registration~\citep{Modat2014reg}. We also used a public test-retest (TRT) dataset~\citep{maclarenReliabilityBrainVolume2014} that contains 120 T1w MRI scans (three subjects, each scanned twice for 20 consecutive days) to assess the in-session test-retest reliability of our model. 

\subsection{Image and surface representation}
Throughout this work, we represent an MR image as a three-dimensional tensor $X\in\mathbb{R}^{H\times W \times D}$ of height $H$, width $W$, and depth $D$, containing a single channel of gray values. A set of $K_i + 1$ MR images from a certain subject $i$ is given as an ordered sequence $\{X_{i,j} \, | \, j=0,\ldots, \, K_i\}$. {\rev After the initial scan $X_{i,0}$, which does not have a particular role and is simply the first scan in the sequence, each subject can have an arbitrary number of $K_i\geq1$ follow-up scans.} Slightly abusing nomenclature, we will also refer to scans from certain visits, i.e., at certain timepoints, simply as \emph{visits} for the sake of compactness.

The tissue boundaries of one cerebral hemisphere, namely the inner WM surface, which divides white and gray matter, and the outer pial surface, which divides gray matter and cerebrospinal fluid, are considered to be two-dimensional manifolds $\mathcal{M} \subset \mathbb{R}^3$ embedded in three-dimensional Euclidean space. We represent the surfaces as closed triangular meshes $\mathcal{M}=\{\mathcal{V}, \mathcal{F}\}$, where each mesh consists of $V$ vertices $\mathcal{V}\in\mathbb{R}^{V \times 3}$ and $F$ faces $\mathcal{F}\in \mathbb{N}_0^{F \times 3}$, storing the indices to vertices. Implicitly, the faces also define a set of edges $\mathcal{E}$ that connect vertices pairwise. From a sequence of visits from subject $i$, we aim to reconstruct corresponding cortical surfaces $\{\mathcal{M}_{i,0} = \{\mathcal{V}_{i,0}, \mathcal{F}_{i,0}\}, \, \ldots, \, \mathcal{M}_{i,K_i}=\{\mathcal{V}_{i,K_i}, \mathcal{F}_{i,K_i}\}\}$, with fixed connectivity, i.e., $\mathcal{F}_{i,0}\equiv\ldots\equiv\mathcal{F}_{i,K_i}$. Although there is no order among faces or vertices \textit{per se}, we adopt an arbitrary but fixed order in our implementation to represent a fixed mesh connectivity and precise anatomical locations in the brain. We denote the within-subject mesh template of subject $i$ as $\mathcal{T}_i$ and population templates simply as $\mathcal{T}$. Slightly abusing the notation, we will refer to the within-subject template meshes and the within-subject template's vertices equally with $\mathcal{T}_i$. The vertices define the mesh entirely, given that the connectivity is pre-defined.

\subsection{Model architecture and implementation} \label{sec:architecture}
Our V2C-Long framework builds upon the V2C-Flow model~\citep{Bongratz2024v2cflow}. 
%, which we describe briefly in the following. 
{\rev Briefly, V2C-Flow learns to deform the generic FsAverage population template \citep{Fischl1999fsaverage} to individual cortical surfaces, thereby establishing correspondences among all reconstructed surfaces and to the template. In contrast to other recent cortical surface reconstruction methods, V2C-Flow incorporates virtual edges between WM and pial surfaces to mitigate their intersection.}
V2C-Flow employs a UNet~\citep{ronnebergerUNetConvolutionalNetworks2015}-like convolutional neural network (CNN) to extract multi-resolution image features from the input image, and it maps them onto the template vertices by trilinear interpolation. From these vertex features, a graph neural network (GNN)~\citep{morris2019} computes a displacement field that is integrated numerically, deforming the template to the brain contours. Formally, the deformation is described by the initial value problem
\begin{equation} \label{eq:v2cflow}
    \frac{d\mathcal{V}(t)}{dt} = f_{\theta}(t, X, \mathcal{V}(t)); \quad \mathcal{V}(0) = \mathcal{T},
\end{equation}
where $X$ is the input image, $\mathcal{T}$ is the population template, and $\theta$ are the parameters of the CNN and GNN. To solve \Cref{eq:v2cflow}, we use an Euler integration scheme with five integration steps ({\rev step size 0.2}) as in the original work. The final reconstruction is given by $\mathcal{V}(1)$.  

In V2C-Long, we employ two consecutive V2C-Flow models for within-subject template creation (cf.~\Cref{sec:creation}) and within-subject template deformation (cf.~\Cref{sec:deformation}). {\rev Each V2C-Flow model is trained jointly with a combination of voxel- (cross-entropy) and mesh-loss (curvature-weighted Chamfer, edge, and normal consistency) functions for WM and pial surfaces at once. The loss functions and their weighting factors are described in detail in \citep{Bongratz2024v2cflow}. During training, we sample 100,000 points randomly per WM/pial surface from the reference meshes, obtained from FreeSurfer~(v7.2), as the target for the curvature-weighted Chamfer loss}. Our implementation is based on Python~(v3.9), PyTorch~(v1.10.0), Cuda~(v11.3), PyTorch3d~(v0.6.1), and automatic mixed precision.
The code and models will be available at \url{https://github.com/ai-med/Vox2Cortex}.

\subsection{Within-subject template creation} \label{sec:creation}

In contrast to existing image-based within-subject template creation~\citep{reuterWithinsubjectTemplateEstimation2012,Li2012consistentreconstruction,Ashburner2013longMRI}, we present a new approach to obtain these templates directly in mesh space, i.e., from the vertices that define the meshes together with the faces. More precisely, we propose to leverage V2C-Flow and, for a certain subject $i$, compute the within-subject template $\mathcal{T}_i$ from the mean vertex locations in a sequence of visits:
\begin{equation} \label{eq:template}
    \mathcal{T}_i = \frac{1}{K_i+1}\sum_{j=0}^{K_i}\mathcal{V}_{i,j}.
\end{equation} 
\begin{theorem} \label{thm:template}
    \Cref{eq:template} solves the initial value problem $\frac{d\mathcal{T}_i(t)}{dt} = \bar{f}_{\theta}(t, X_{\rev 0,1,\ldots,K_i}, \mathcal{T}(t)); \, \mathcal{T}_i(0) = \mathcal{T}$ in a first order approximation, where $\bar{f}$ is the mean deformation field across all of the subject's visits.
\end{theorem}
We prove this statement in \Cref{sec:appendix}. 
\Cref{thm:template} 
{\rev implies that the template we obtain} from \Cref{eq:template} is equivalent to solving \Cref{eq:ode} on the subject level after aggregating all of the subject's flow fields into a joint deformation $\bar{f}$, following the standard definition of deformable anatomical templates~\citep{grenander1998}.
The prerequisite for the mean aggregation is corresponding vertices, which we get from V2C-Flow~\citep{Bongratz2024v2cflow}. {\rev {\revtwo See Supplementary Figures~5 and~6 for an illustration of vertex correspondences in V2C-Flow/V2C-Long.}}
Since the sample mean is an unbiased estimator, the within-subject templates obtained from \Cref{eq:template} are not biased toward any visit. Nonetheless, including all available visits in the aggregation is vital to avoid asymmetries, e.g., toward earlier visits if later ones were omitted.

To enrich the within-subject templates with additional information about an individual's {\rev cortical geometry}, we concatenate the vertex features extracted by the two graph neural network (GNN) blocks in V2C-Flow (obtained right before the output layers that compute the deformation) to the vertex coordinates. {\rev These features (of dimension $64\times2$) contain geometric information beyond bare vertex coordinates as they guide the mesh flow.} 
Formally, we compute the mean in \Cref{eq:template} from the resulting generalized vertices, which encompass both the spatial coordinates and the associated vertex features.

\subsection{Within-subject template deformation} \label{sec:deformation}
In a second step, we deform the within-subject template again to the contours of each visit to improve the temporal consistency of reconstructed surfaces compared to the initial reconstruction.  
To this end, we train a second V2C-Flow model 
with subject-specific input templates, i.e., 
\begin{equation} \label{eq:ode}
    \frac{d\mathcal{V}_{i,j}(t)}{dt} = f_\theta(t,{\rev X_{i,j}}, \mathcal{V}_{i,j}(t)); \quad \mathcal{V}_{i,j}(0) = \mathcal{T}_i.
\end{equation}
Importantly, we provide the template only as a starting point, but the temporal deformation is not constrained further to avoid over-regularization~\citep{reuterAvoidingAsymmetryinducedBias2011}. To speed up the training process in this second stage of V2C-Long, we provide the weights of the previously trained template-creation model as an initialization (except for input layers due to the additional vertex features).

\subsection{Evaluation metrics} \label{sec:metrics}

We use two sets of metrics to evaluate the quality and consistency of the predicted meshes. We provide the respective definitions in the following.

\subsubsection{Surface quality metrics}

The average symmetric surface distance (ASSD) and symmetric Hausdorff distance (HD) measure the concordance of two contours in average (ASSD) and in the worst case (HD). 
{\rev For better robustness, we use a percentile-based version of the HD (HD$_{\text{X}}$).
To compute distances between surfaces, we sample points uniformly from the predicted and reference meshes, $\mathcal{M}^{\text{Pred}}$ and $\mathcal{M}^{\text{Ref}}$, respectively,
yielding point sets $\mathcal{P}^{\text{Pred}}$ and $\mathcal{P}^{\text{Ref}}$. 
From the point sets and the meshes, we can compute point-to-surface distances in a bilateral manner. As a measure of topological accuracy, we also compute the number of self-intersecting faces (SIF) from the predicted meshes.}

\noindent
\textbf{ASSD.}
Formally, the ASSD computes as

\begin{equation}
  \mathrm{ASSD}(\mathcal{M}^{\text{Pred}}, \mathcal{M}^{\text{\rev Ref}}) = \frac{\sum_{p \in \mathcal{P}^{\text{Pred}}} d(p,\mathcal{M}^{\text{Ref}}) + \sum_{p \in \mathcal{P}^{\text{Ref}}} d(p,\mathcal{M}^{\text{Pred}})}{\abs{\mathcal{P}^{\text{Pred}}} + \abs{\mathcal{P}^{\text{Ref}}}},
\end{equation}

where $d(p, \mathcal{M})$ is the Euclidean point-to-face distance between a point $x$ and the closest point in the mesh $\mathcal{M}$. In our implementation, we use $\abs{\mathcal{P}^{\text{Pred}}} = \abs{\mathcal{P}^{\text{Ref}}} = 100,000$.

\textbf{HD$_{\text{X}}$.} Similarly, the symmetric HD computes as 

\begin{equation}
  \mathrm{HD}_{\text{X}}(A,B) = \max\bigl(\mathrm{Q}_{\text{X}}(\{d(p,\mathcal{M}^{\text{Pred}}) \mid p \in \mathcal{M}^{\text{Ref}}\}), \mathrm{Q}_{X}(\{d(p,\mathcal{M}^{\text{Ref}}) \mid p \in \mathcal{M}^{\text{Pred}}\})\bigr).
\end{equation}
To make the HD robust to outliers, a percentile Q$_{\text{X}}$ can be selected (HD$_{100}$ corresponds to the standard Hausdorff distance which is easily dominated by a single vertex).

\noindent
\textbf{SIF.}
Ideally, cortical surfaces should be represented by watertight 2-manifolds, i.e., closed surfaces without holes, handles, or self-intersections. While the template-deformation approach in V2C-Long circumvents holes and handles entirely, self-intersections can still occur. We compute the number of self-intersecting faces (SIF), i.e., the ratio of faces intersecting with another face from the same mesh, using PyMeshLab~(v2022.2).

\subsubsection{Surface consistency metrics}

We use the following metrics to evaluate the test-retest reliability, i.e., the consistency of reconstructions from images acquired within a short time, and the longitudinal consistency, i.e., the consistency of reconstructions with longer time periods between the visits.

\noindent
\textbf{MCVar \& CThVar.}
We quantify point-wise longitudinal consistency via the variance in derived morphological descriptors, namely cortical curvature (mean curvature variance, MCVar)~\citep{meyer2003} and thickness (cortical thickness variance, CThVar)~\citep{Fischl2000thickness}. 
The rationale is that local variations in curvature (the cortical sheet is tightly folded) and thickness are usually greater than anatomical alterations over time~\citep{Fischl2000thickness}. Hence, the variance in these measures should be smaller with better alignment (albeit not necessarily zero). 
Formally, we compute the unbiased sample variance
\begin{equation}
    \text{MCVar}(v) = \frac{1}{K_i} \sum_{j=0}^{K_i}  (\kappa_{j} - \bar{\kappa})^2, \quad \bar{\kappa} = \frac{1}{K_i + 1} \sum_{j=0}^{K_i} \kappa_{j},  
\end{equation}
where $\kappa$ is the discrete mean curvature associated with a vertex $v\in\mathcal{T}_i$. The CThVar is defined analogously. Note that the curvature can be computed individually for WM and pial surfaces, whereas only a single CThVar value is obtained from both surfaces. We compute a robust score on the subject level by taking the median across all vertices. {\rev For estimating cortical thickness, we follow the bilateral method proposed by \citet{Fischl2000thickness}. We calculate the shortest distance from a vertex on the WM surface to the pial surface, and vice versa, then take the average of these distances at each vertex.}

\noindent
\textbf{ParcF1.}
To further assess the longitudinal consistency on the region level, we map the Destrieux atlas~\citep{destrieuxAutomaticParcellationHuman2010}, a fine-grained cortical atlas dividing the cortex into gyri and sulci, from FreeSurfer's FsAverage to the predicted surfaces (assigning a certain class to each vertex). Given the respective class labels, we measure the overlap, i.e., the consistency, of a region in the longitudinal sequence by pairwise comparison of nearest-neighbor vertex classes. On the surface level, the F1 score is weighted according to the different region sizes.

\subsection{Vertex-wise linear mixed effects regression} \label{sec:mle}
To compare the pattern of longitudinal cortical thickness in two diagnostic groups,
we employ a vertex-wise linear mixed effects (LME) regression model~\citep{bernal-rusielStatisticalAnalysisLongitudinal2013,Wachinger2016wholebrainasymmetry}. This model is defined as
\begin{equation}
    \text{CTh}_{i,j} = \beta_0 + \beta_1 B_i + \beta_2 W_{i,j} + \beta_3 D_i + b_{0,i} + b_{1,i} W_{i,j},
\end{equation}
where, for subject $i$, the age at {\rev initial visit $B_i$, the time $W_{i,j}$ from initial to follow-up visit $j$}, and the stable diagnosis $D_i$ is considered. It allows for individual intercepts and slopes via random effects regression coefficients $b_{0,i}$, $b_{1,i}$. The fixed, non-individual regression coefficients $\beta_0, \beta_1, \beta_2$ model the global effects on cortical thickness. We use the LME implementation in Statsmodels (v0.14.1).

\section{Results}

In this section, we report results in terms of longitudinal consistency (\Cref{sec:rec-consistency}), accuracy (\Cref{sec:rec-accuracy}), and in-session test-retest reliability (\Cref{sec:trt}) using the metrics described in \Cref{sec:metrics}. In addition, we provide an ablation study of V2C-Long in \Cref{sec:ablation}. Finally, we replicate differences in longitudinal cortical thickness between patients diagnosed with Alzheimer's disease and a healthy control population using the LME regression model described in \Cref{sec:mle}. 

\subsection{Experimental setting}

We compare V2C-Long to the longitudinal FreeSurfer processing (FS-Long)~\citep{reuterWithinsubjectTemplateEstimation2012}. Additionally, we include recent deep learning-based cortex reconstruction methods. Namely, we implemented TopoFit\footnote{We adapted TopoFit for pial surfaces (originally only for WM surfaces).}~\citep{hoopesTopoFitRapidReconstruction}, CF$^{++}$\footnote{We used the CF$^{++}$ template with $\approx$140,000 vertices for best comparison to the other methods.}~\citep{cruzCorticalFlowBoostingCortical2022}, V2C-Flow~\citep{Bongratz2024v2cflow}, and V2CC~\citep{rickmannVertexCorrespondenceCortical2023} consistently for the right hemisphere based on the original repositories. {\rev Although these methods are not optimized for longitudinal data as thoroughly as FS-Long and V2C-Long, they are crucial baselines due to their proven accuracy and architectural similarity.} {\revtwo We used the raw output of all reconstruction methods without further surface-based post-processing, as this is most comparable to V2C-Long. Nevertheless, we also consider spherical registration-based post-processing as another baseline. Specifically, we applied the spherical registration implemented in FreeSurfer~(FS-Reg)~\citep{fischlFreeSurfer2012}, with FsAverage as the registration target and barycentric interpolation of vertex coordinates, to V2C-Flow surfaces. We refer to this approach as V2C-Flow/FS-Reg.} Results for a V2C-Long model trained on both hemispheres are in the Supplementary Material. We trained all deep-learning methods in the same setting using our longitudinal ADNI training set and Nvidia A100 GPUs with 40GB VRAM. We consistently selected the model with the lowest reconstruction error (ASSD) on the validation set.

\subsection{Longitudinal consistency} \label{sec:rec-consistency}

\begin{table}[t]
\renewcommand\bfdefault{b}% rather than bx
\begin{threeparttable}
\setlength{\tabcolsep}{4pt}

\centering
\caption{Consistency metrics by surface and method for the ADNI and OASIS test sets. Values are mean{\scriptsize$\pm$SD} over all subjects in the respective dataset and the best results are \textbf{highlighted}.}
\label{tab:adni_correspondence}
\begin{tabular}{@{}llccccc@{}}
\toprule
\multicolumn{1}{c}{} & \multicolumn{1}{c}{} &  \multicolumn{2}{c}{WM surface} &  \multicolumn{2}{c}{Pial surface} &  \multicolumn{1}{c}{Both}\\
\cmidrule(lr){3-4}
\cmidrule(lr){3-4}
\cmidrule(lr){5-6}
\cmidrule(lr){5-6}
\cmidrule(lr){7-7}
\cmidrule(lr){7-7}
& Model & MCVar$\downarrow$ & ParcF1$\uparrow$ & MCVar$\downarrow$ & ParcF1$\uparrow$ & CThVar$\downarrow$ \\
\toprule
\multirow{6}{*}{\rotatebox[origin=c]{90}{ADNI}} & V2C-Long & \textbf{0.020}\scriptsize\ensuremath{\pm}0.020 & \textbf{0.971}\scriptsize\ensuremath{\pm}0.015 & \textbf{0.012}\scriptsize\ensuremath{\pm}0.015 & \textbf{0.966}\scriptsize\ensuremath{\pm}0.017 & \textbf{0.016}\scriptsize\ensuremath{\pm}0.010 \\
&V2C-Flow~\citep{Bongratz2024v2cflow} & 0.060\scriptsize\ensuremath{\pm}0.023 & 0.924\scriptsize\ensuremath{\pm}0.030 & 0.047\scriptsize\ensuremath{\pm}0.018 & 0.920\scriptsize\ensuremath{\pm}0.028 & 0.038\scriptsize\ensuremath{\pm}0.027 \\

%/mnt/nas/Projects/V2C-long/correspondence/v2c_flow_seq_91_writable/test_template_fsaverage-smooth-rh_ADNI_long_FS_cross_n_5
& \revtwo V2C-F./FS-Reg & 
\revtwo 0.039\scriptsize\ensuremath{\pm}0.014 & 
\revtwo 0.934\scriptsize\ensuremath{\pm}0.053 & 
\revtwo 0.033\scriptsize\ensuremath{\pm}0.012 & 
\revtwo 0.927\scriptsize\ensuremath{\pm}0.051 &
\revtwo 0.028\scriptsize\ensuremath{\pm}0.029
\\

&V2CC~\citep{rickmannVertexCorrespondenceCortical2023} & 0.032\scriptsize\ensuremath{\pm}0.009 & 0.958\scriptsize\ensuremath{\pm}0.016 & 0.028\scriptsize\ensuremath{\pm}0.009 & 0.945\scriptsize\ensuremath{\pm}0.018 & 0.021\scriptsize\ensuremath{\pm}0.009 \\
&CF$^{++}$~\citep{cruzCorticalFlowBoostingCortical2022}\tnote{$a$} & 0.046\scriptsize\ensuremath{\pm}0.017 & -- & 0.046\scriptsize\ensuremath{\pm}0.017 & -- & 0.034\scriptsize\ensuremath{\pm}0.027 \\
&TopoFit~\citep{hoopesTopoFitRapidReconstruction} & 0.049\scriptsize\ensuremath{\pm}0.018 & 0.912\scriptsize\ensuremath{\pm}0.050 & 0.050\scriptsize\ensuremath{\pm}0.020 & 0.903\scriptsize\ensuremath{\pm}0.050 & 0.031\scriptsize\ensuremath{\pm}0.014 \\
&FS-Long~\citep{reuterWithinsubjectTemplateEstimation2012} & 0.032\scriptsize\ensuremath{\pm}0.013 & 0.968\scriptsize\ensuremath{\pm}0.050 & 0.031\scriptsize\ensuremath{\pm}0.015 & 0.953\scriptsize\ensuremath{\pm}0.050 & 0.032\scriptsize\ensuremath{\pm}0.019 \\

\midrule

\multirow{6}{*}{\rotatebox[origin=c]{90}{OASIS}} & V2C-Long & \textbf{0.021}\scriptsize\ensuremath{\pm}0.007 & 0.963\scriptsize\ensuremath{\pm}0.016 & \textbf{0.013}\scriptsize\ensuremath{\pm}0.005 & \textbf{0.953}\scriptsize\ensuremath{\pm}0.019 & \textbf{0.016}\scriptsize\ensuremath{\pm}0.007 \\
 & V2C-Flow~\citep{Bongratz2024v2cflow}  & 0.064\scriptsize\ensuremath{\pm}0.021 & 0.911\scriptsize\ensuremath{\pm}0.020 & 0.050\scriptsize\ensuremath{\pm}0.017 & 0.904\scriptsize\ensuremath{\pm}0.022 & 0.036\scriptsize\ensuremath{\pm}0.015 \\

 & \revtwo V2C-F./FS-Reg &
 \revtwo 0.038\scriptsize\ensuremath{\pm}0.012 & 
 \revtwo 0.929\scriptsize\ensuremath{\pm}0.018 & 
 \revtwo 0.033\scriptsize\ensuremath{\pm}0.011 & 
 \revtwo 0.917\scriptsize\ensuremath{\pm}0.020 & 
 \revtwo 0.025\scriptsize\ensuremath{\pm}0.011 
 \\
 
 & V2CC~\citep{rickmannVertexCorrespondenceCortical2023}  & 0.035\scriptsize\ensuremath{\pm}0.010 & 0.947\scriptsize\ensuremath{\pm}0.017 & 0.034\scriptsize\ensuremath{\pm}0.011 & 0.929\scriptsize\ensuremath{\pm}0.021 & 0.023\scriptsize\ensuremath{\pm}0.010 \\
 & CF$^{++}$~\citep{cruzCorticalFlowBoostingCortical2022}\tnote{$a$} & 0.050\scriptsize\ensuremath{\pm}0.018 & -- & 0.051\scriptsize\ensuremath{\pm}0.019 & -- & 0.037\scriptsize\ensuremath{\pm}0.018 \\
 & TopoFit~\citep{hoopesTopoFitRapidReconstruction} & 0.060\scriptsize\ensuremath{\pm}0.019 & 0.885\scriptsize\ensuremath{\pm}0.023 & 0.067\scriptsize\ensuremath{\pm}0.024 & 0.873\scriptsize\ensuremath{\pm}0.025 & 0.036\scriptsize\ensuremath{\pm}0.014 \\
 & FS-Long~\citep{reuterWithinsubjectTemplateEstimation2012} & 0.033\scriptsize\ensuremath{\pm}0.013 & \textbf{0.964}\scriptsize\ensuremath{\pm}0.017 & 0.040\scriptsize\ensuremath{\pm}0.020 & 0.941\scriptsize\ensuremath{\pm}0.021 & 0.030\scriptsize\ensuremath{\pm}0.013 \\

\bottomrule
\end{tabular}

\begin{tablenotes}[flushleft]\footnotesize
\item[${a}$]Computation of the ParcF1 score is not possible for CF$^{++}$ due to the custom template for which no atlas is available.
\end{tablenotes}

\end{threeparttable}
\end{table}

\begin{figure}
    \centering
    \includegraphics[width=0.8\textwidth]{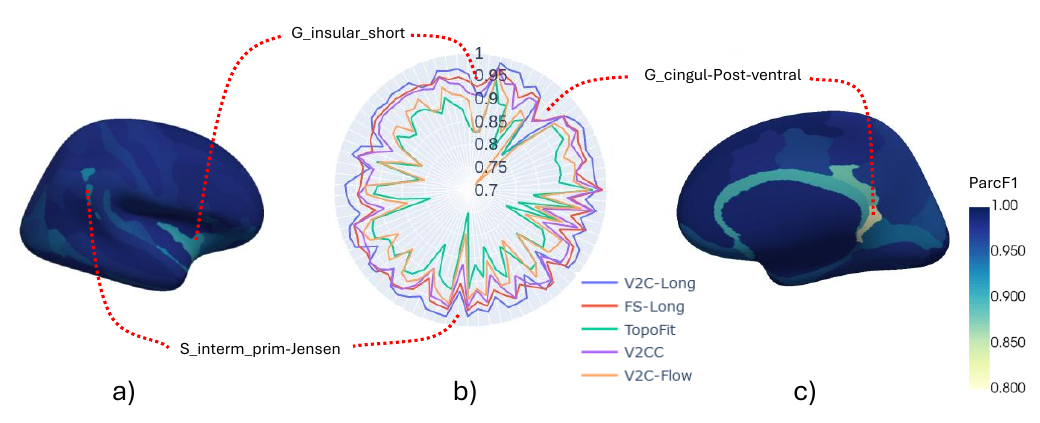}
    \caption{Region-wise consistency of reconstructed surfaces based on the Destrieux atlas (mean of WM and pial surfaces, ADNI test set). a) Lateral view of per-region ParcF1 scores (higher is better) from V2C-Long. b) Comparison of ParcF1 scores for all 75 regions from different methods. c) {\rev Medial view} of per-region ParcF1 scores from V2C-Long. We also indicate the location and name of {\rev three regions} in the atlas and the polar chart. A list of all Destrieux regions in the order of the plot (counter-clockwise) is in the Supplementary Material. For plotting, we used Pyvista~(v0.35.2) and Plotly~(v5.22.0).}
    \label{fig:f1parc}
\end{figure}

\begin{figure}
    \centering
    \includegraphics[width=0.9\textwidth]{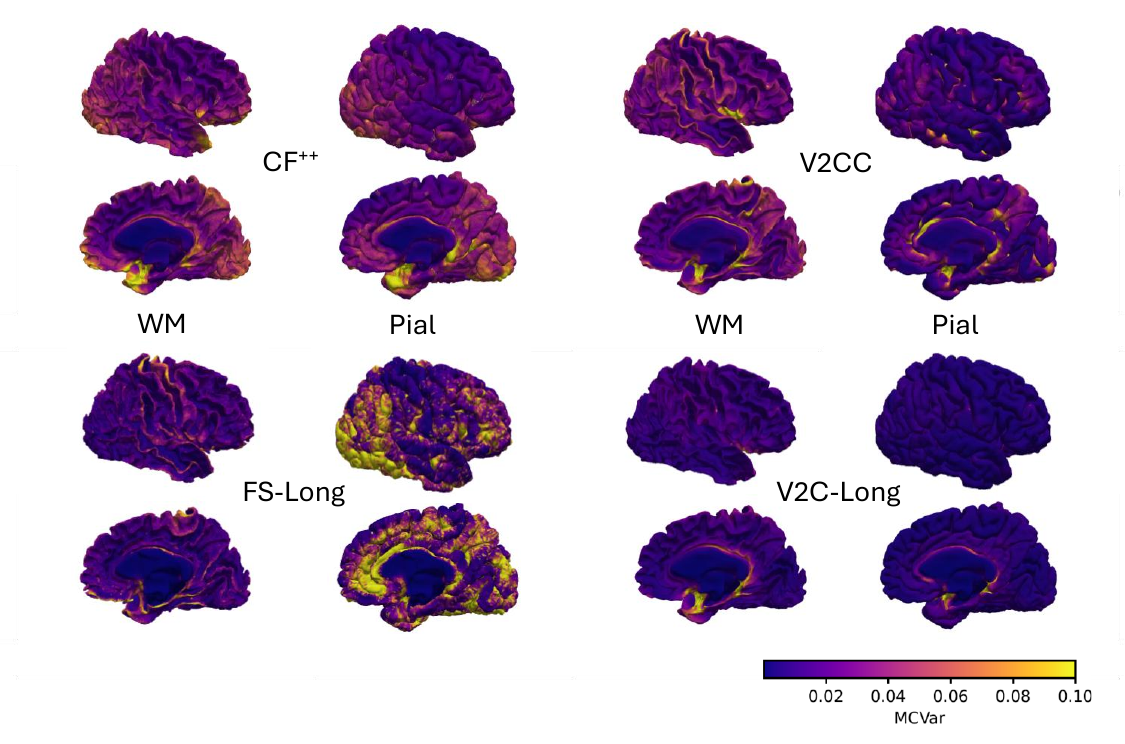}
    \caption{\rev Vertex-wise longitudinal mean curvature variance (MCVar, lower is better) for CF$^{++}$, V2CC, FS-Long, and V2C-Long. We show the mean over all subjects in our ADNI test set from lateral and {\rev medial views}. We use the same exemplary individual anatomy for each plot to avoid visual differences due to the different templates in CF$^{++}$ and the other methods.
    }
    \label{fig:mcvar}
\end{figure}

\begin{figure}
\centering
\begin{adjustbox}{center}
    \includegraphics[width=\textwidth]{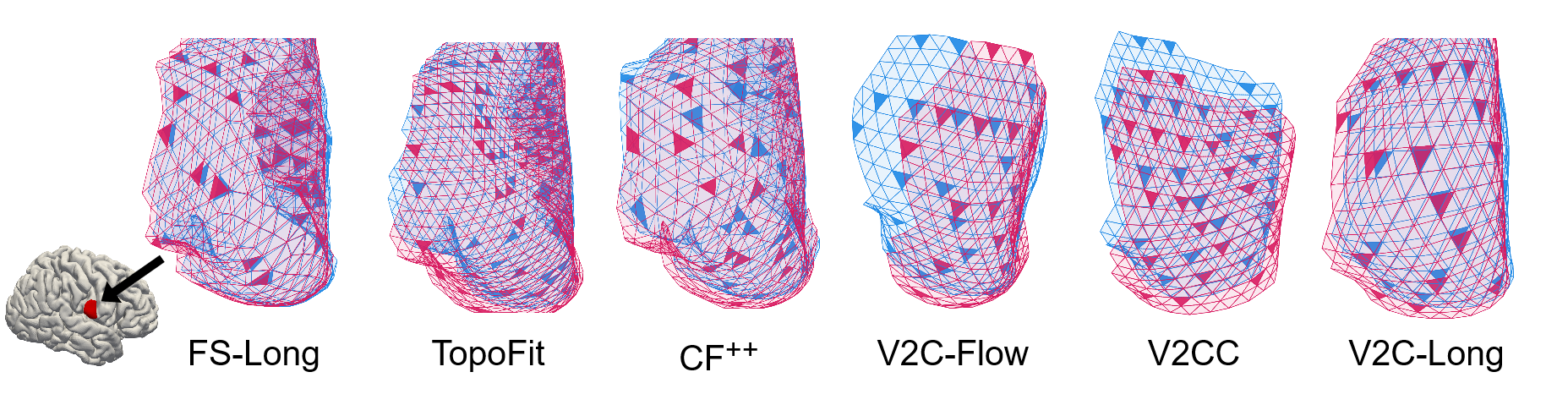}
\end{adjustbox}
\caption{{\rev Cortical surfaces from longitudinal methods (FS-Long and V2C-Long) are consistent across time --- a property fulfilled by none of the recent cross-sectional methods (TopoFit, CF$^{++}$, V2C-Flow, V2CC). We depict a right-hemisphere region of reconstructed pial surfaces from two visits (red and blue, respectively) of a subject in our ADNI test set.} Ideally, colored triangles match.}
\label{fig:mesh_correspondence}
\end{figure}

\Cref{tab:adni_correspondence} shows the longitudinal consistency metrics of all implemented methods for inner (WM) and outer (pial) cortical surfaces. {\revtwo Corresponding plots are in Supplementary Figure~4}. We observe a substantial improvement with V2C-Long over all baseline methods --- including FS-Long on the internal (ADNI) and external (OASIS) test sets. The variance in discrete mean curvature (MCVar) is reduced by up to 61\% (pial, ADNI) compared to the closest competitor, V2CC, and by up to 67\% (pial, OASIS) compared to FS-Long. Regarding cortical thickness, V2C-Long achieves an improvement of 24\% (ADNI) and 30\% (OASIS) over the best alternative method. On the region level (ParcF1), V2C-Long is again at the forefront of the best methods, only beaten by FS-Long on OASIS WM surfaces by 0.001. 
{
\revtwo The spherical registration improves the longitudinal consistency compared to the raw output of V2C-Flow. However, V2C-Long remains significantly superior across both datasets.
}

\Cref{fig:f1parc} shows the longitudinal consistency for individual Destrieux atlas regions. In the polar chart, each angle corresponds to a certain region, and the radial distance represents its F1 score, i.e., the longitudinal consistency. For V2C-Long, we plot the region-wise consistency on inflated brain surfaces. We connect three of them, i.e., sulcus
intermedius primus of Jensen (S\_interm\_prim-Jensen), posterior-ventral cingulate gyrus (G\_cingul-Post-ventral), and short insular gyri (G\_insular\_short) to the corresponding kinks in the polar chart. These regions are either comparably small (S\_interm\_prim-Jensen), located in the medial area of the cortex (G\_cingul-Post-ventral), or deep in the insular cortex (G\_insular\_short), which makes them particularly difficult to reconstruct for all implemented methods. The polar chart further reveals that the consistency in V2C-Long is superior to all other methods for most (62/75) of the sulcal and gyral regions, followed by FS-Long (13/75). 

In \Cref{fig:mcvar}, we plot the MCVar per vertex for CF$^{++}$, V2CC, FS-Long, and V2C-Long in average over the ADNI test set. {\rev Note that FS-Long curvature measures need to be registered and re-sampled for vertex-wise comparison across subjects. All other methods directly permit averaging per-vertex values on the group level. We observe that V2C-Long achieves high consistency (i.e., low MCVar) across the entire cortex and prevents inconsistencies in the occipital lobe produced by CF$^{++}$ and FS-Long. V2CC and FS-Long further have considerable longitudinal inconsistencies in pre- and postcentral gyri (WM surface), again alleviated by V2C-Long. Inconsistencies around the cingulate cortex and the {\revtwo medial wall~(labeled as ``unknown'' in the Desikan-Killany atlas~\citep{Desikan2006})}, stretching in parts into parahippocampal, entorhinal, and insular areas, are highlighted in all methods. Yet, they are less pronounced in V2C-Long, especially on the pial surface.}

\Cref{fig:mesh_correspondence} shows superimposed and color-coded pial surfaces from two visits of an individual from our ADNI test set. {\rev We recognize that, qualitatively, FS-Long and V2C-Long are the only methods that match individual triangles from the two different visits.
The cross-sectional methods (TopoFit, CF$^{++}$, V2C-Flow, and V2CC), on the other hand, failed to provide qualitatively consistent meshes across time.}

\subsection{Test-retest reliability} \label{sec:trt}
\begin{table}[tbh]
\setlength{\tabcolsep}{4pt}
\renewcommand\bfdefault{b}% rather than bx
\centering
\caption{Test-retest reliability of V2C-Long compared to V2CC and FS-Long. Values are mean{\scriptsize\ensuremath{\pm}SD} over the three subjects in the TRT dataset. The best results are \textbf{highlighted}.}

\begin{tabular}{@{}llccccc@{}}
\toprule
& \multicolumn{1}{c}{} &  \multicolumn{2}{c}{WM surface} &  \multicolumn{2}{c}{Pial surface} & \multicolumn{1}{c}{Both}\\
\cmidrule(lr){3-4}
\cmidrule(lr){3-4}
\cmidrule(lr){5-6}
\cmidrule(lr){5-6}
\cmidrule(lr){7-7}
& Method 
& MCVar$\downarrow$
& ParcF1$\uparrow$ 
& MCVar$\downarrow$
& ParcF1$\uparrow$ 
& CThVar$\downarrow$
\\

\midrule

\parbox[t]{3mm}{\multirow{3}{*}{\rotatebox[origin=c]{90}{TRT}}}

       & V2C-Long  
       & \textbf{0.020}\scriptsize\ensuremath{\pm}0.002
       & 0.982\scriptsize\ensuremath{\pm}0.001
    
       & \textbf{0.017}\scriptsize\ensuremath{\pm}0.001
       & \textbf{0.966}\scriptsize\ensuremath{\pm}0.003
       & \textbf{0.020}\scriptsize\ensuremath{\pm}0.001
       \\

       & V2CC~\citep{rickmannVertexCorrespondenceCortical2023} 
        & 0.040\scriptsize\ensuremath{\pm}0.006
        & 0.966\scriptsize\ensuremath{\pm}0.004
        & 0.043\scriptsize\ensuremath{\pm}0.008
        & 0.947\scriptsize\ensuremath{\pm}0.002
        & 0.031\scriptsize\ensuremath{\pm}0.003
        
        \\
        
      & FS-Long~\citep{reuterWithinsubjectTemplateEstimation2012}  
     & 0.041\scriptsize\ensuremath{\pm}0.004
       & \textbf{0.983}\scriptsize\ensuremath{\pm}0.001
       & 0.123\scriptsize\ensuremath{\pm}0.023
       & 0.947\scriptsize\ensuremath{\pm}0.004
       & 0.036\scriptsize\ensuremath{\pm}0.000
       \\

       \bottomrule
    \end{tabular}
    \label{tab:trt}
\end{table}

In \Cref{tab:trt}, we validate the in-session test-retest reliability, i.e., the consistency of reconstructions from images acquired within a short time and in a highly consistent setting, using the TRT dataset. In contrast to the evaluation based on ADNI and OASIS in the previous section, the number of subjects in the TRT dataset is much smaller ($n=3$), and the number of visits is much higher (40 per subject).
Here, we limit the comparison to the best methods from \Cref{tab:adni_correspondence}, i.e., V2C-Long, V2CC, and FS-Long. The results are largely consistent with the observations on ADNI and OASIS despite the different nature of the TRT data. V2C-Long achieves the best scores in all metrics except for the parcellation consistency (ParcF1) on WM surfaces, where FS-Long is again superior by 0.001. The largest improvement is achieved for the curvature variance (MCVar) on WM surfaces, where V2C-Long reduces the inconsistencies by half compared to V2CC and FS-Long.

\subsection{Reconstruction accuracy \rev and diagnostic differentiation} \label{sec:rec-accuracy}

\begin{table}[t]
\renewcommand\bfdefault{b}% rather than bx

\centering
\caption{Reconstruction metrics by surface and method for the ADNI and OASIS test sets. Values are mean{\scriptsize$\pm$SD} in mm over all scans in the respective dataset, and the best non-FreeSurfer results are \textbf{highlighted}.}
\begin{adjustbox}{center}
\begin{tabular}{@{}llcccccc@{}}
\toprule
\multicolumn{1}{c}{} & \multicolumn{1}{c}{} &  \multicolumn{3}{c}{WM surface} &  \multicolumn{3}{c}{Pial surface}\\
\cmidrule(lr){3-5}
\cmidrule(lr){3-5}
\cmidrule(lr){6-8}
\cmidrule(lr){6-8}
&Model & ASSD$\downarrow$ & $\textsf{HD}_{90}$$\downarrow$ & $\textsf{HD}_{99}$$\downarrow$ & ASSD$\downarrow$ & $\textsf{HD}_{90}$$\downarrow$ & $\textsf{HD}_{99}$$\downarrow$ \\
\midrule
\parbox[t]{3mm}{\multirow{6}{*}{\rotatebox[origin=c]{90}{ADNI}}}
&V2C-Long 
& \textbf{0.177}\scriptsize\ensuremath{\pm}0.161 
& \textbf{0.410}\scriptsize\ensuremath{\pm}0.756 
& \textbf{1.115}\scriptsize\ensuremath{\pm}1.329 
& \textbf{0.174}\scriptsize\ensuremath{\pm}0.161 
& \textbf{0.409}\scriptsize\ensuremath{\pm}0.731 
& 1.374\scriptsize\ensuremath{\pm}1.308 \\

&V2C-Flow 
& 0.186\scriptsize\ensuremath{\pm}0.116 
& 0.427\scriptsize\ensuremath{\pm}0.644 
& 1.168\scriptsize\ensuremath{\pm}1.429 
& 0.180\scriptsize\ensuremath{\pm}0.109 
& 0.419\scriptsize\ensuremath{\pm}0.566 
& 1.392\scriptsize\ensuremath{\pm}1.342 \\

&\revtwo V2C-F./FS-Reg
&\revtwo 0.203\scriptsize\ensuremath{\pm}0.137
&\revtwo 0.458\scriptsize\ensuremath{\pm}0.705
&\revtwo 1.214\scriptsize\ensuremath{\pm}1.298
&\revtwo 0.204\scriptsize\ensuremath{\pm}0.123
&\revtwo 0.471\scriptsize\ensuremath{\pm}0.659
&\revtwo 1.520\scriptsize\ensuremath{\pm}1.237
\\

&V2CC 
& 0.220\scriptsize\ensuremath{\pm}0.036 
& 0.492\scriptsize\ensuremath{\pm}0.089 
& 1.469\scriptsize\ensuremath{\pm}0.465 
& 0.243\scriptsize\ensuremath{\pm}0.040
 & 0.558\scriptsize\ensuremath{\pm}0.104 
 & 1.798\scriptsize\ensuremath{\pm}0.431 \\
 
&CF$^{++}$
& 0.214\scriptsize\ensuremath{\pm}0.140 
& 0.493\scriptsize\ensuremath{\pm}0.812 
& 1.262\scriptsize\ensuremath{\pm}1.467 
& 0.191\scriptsize\ensuremath{\pm}0.129 
& 0.445\scriptsize\ensuremath{\pm}0.802 
& \textbf{1.343}\scriptsize\ensuremath{\pm}1.419 \\

&TopoFit 
& 0.194\scriptsize\ensuremath{\pm}0.033 
& 0.440\scriptsize\ensuremath{\pm}0.088 
& 1.252\scriptsize\ensuremath{\pm}0.389 
& 0.211\scriptsize\ensuremath{\pm}0.038 
& 0.459\scriptsize\ensuremath{\pm}0.087 
& 1.501\scriptsize\ensuremath{\pm}0.408 \\

&\color{gray}FS-Long
& \color{gray}0.151\scriptsize\ensuremath{\pm}0.089 
& \color{gray}0.317\scriptsize\ensuremath{\pm}0.187 
& \color{gray}0.970\scriptsize\ensuremath{\pm}0.517 
& \color{gray}0.145\scriptsize\ensuremath{\pm}0.078 
& \color{gray}0.300\scriptsize\ensuremath{\pm}0.221 
& \color{gray}1.134\scriptsize\ensuremath{\pm}0.550 \\

\midrule

\multirow{6}{*}{\rotatebox[origin=c]{90}{OASIS}} 
&V2C-Long 
& \textbf{0.176}\scriptsize\ensuremath{\pm}0.023 
& \textbf{0.403}\scriptsize\ensuremath{\pm}0.053 
& \textbf{1.171}\scriptsize\ensuremath{\pm}0.363 
& \textbf{0.186}\scriptsize\ensuremath{\pm}0.025 
& \textbf{0.430}\scriptsize\ensuremath{\pm}0.065 
& 1.450\scriptsize\ensuremath{\pm}0.341 \\

 & V2C-Flow 
 & 0.184\scriptsize\ensuremath{\pm}0.024 
 & 0.419\scriptsize\ensuremath{\pm}0.055 
 & 1.212\scriptsize\ensuremath{\pm}0.359 
 & 0.196\scriptsize\ensuremath{\pm}0.024 
 & 0.452\scriptsize\ensuremath{\pm}0.062 
 & 1.494\scriptsize\ensuremath{\pm}0.338 \\

  & \revtwo V2C-F./FS-Reg 
  & \revtwo 0.205\scriptsize\ensuremath{\pm}0.036
  & \revtwo 0.440\scriptsize\ensuremath{\pm}0.073 
  & \revtwo 1.270\scriptsize\ensuremath{\pm}0.492 
  & \revtwo 0.214\scriptsize\ensuremath{\pm}0.024 
  & \revtwo 0.482\scriptsize\ensuremath{\pm}0.066
  & \revtwo 1.573\scriptsize\ensuremath{\pm}0.380 
 \\
 
 & V2CC 
 & 0.222\scriptsize\ensuremath{\pm}0.036 
 & 0.506\scriptsize\ensuremath{\pm}0.082 
 & 1.542\scriptsize\ensuremath{\pm}0.460 
 & 0.282\scriptsize\ensuremath{\pm}0.044 
 & 0.654\scriptsize\ensuremath{\pm}0.116 
 & 2.009\scriptsize\ensuremath{\pm}0.443 \\
 
 & CF$^{++}$
 & 0.226\scriptsize\ensuremath{\pm}0.032 
 & 0.522\scriptsize\ensuremath{\pm}0.078 
 & 1.463\scriptsize\ensuremath{\pm}0.592 
 & 0.210\scriptsize\ensuremath{\pm}0.041 
 & 0.473\scriptsize\ensuremath{\pm}0.078 
 & \textbf{1.401}\scriptsize\ensuremath{\pm}0.359 \\
 
 & TopoFit 
 & 0.197\scriptsize\ensuremath{\pm}0.026 
 & 0.447\scriptsize\ensuremath{\pm}0.065 
 & 1.302\scriptsize\ensuremath{\pm}0.370 
 & 0.228\scriptsize\ensuremath{\pm}0.036 
 & 0.502\scriptsize\ensuremath{\pm}0.086 
 & 1.597\scriptsize\ensuremath{\pm}0.368 \\
 
 & \color{gray}FS-Long 
 & \color{gray}0.126\scriptsize\ensuremath{\pm}0.024 
 & \color{gray}0.264\scriptsize\ensuremath{\pm}0.053 
 & \color{gray}0.917\scriptsize\ensuremath{\pm}0.459 
 & \color{gray}0.132\scriptsize\ensuremath{\pm}0.025
 & \color{gray}0.274\scriptsize\ensuremath{\pm}0.055 
 & \color{gray}1.120\scriptsize\ensuremath{\pm}0.367 \\
\bottomrule
\end{tabular}

\end{adjustbox}
\\
\vspace{0.1cm}

\label{tab:adni_reconstruction}
\end{table}

\begin{figure}[t]
    \centering
    \includegraphics[width=\textwidth]{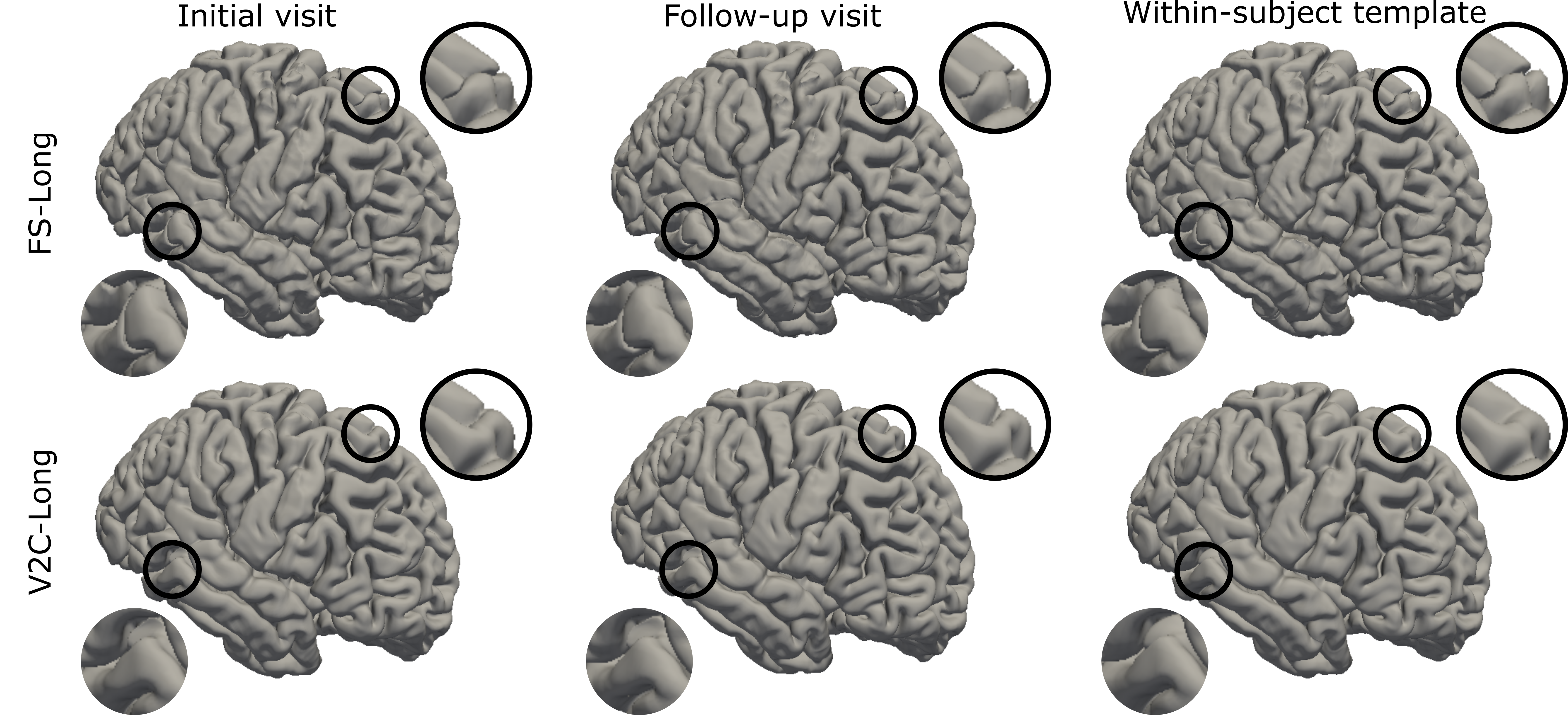}
    \caption{Qualitative comparison of reconstructed pial surfaces of the right hemisphere from a subject of our ADNI test set with two visits. The top row shows the output from FS-Long (v7.2), and the bottom row shows the output from V2C-Long.}
    \label{fig:qualitative}
\end{figure}

In \Cref{tab:adni_reconstruction}, we report the reconstruction accuracy of all implemented methods. {\rev We compute the ASSD, HD$_{90}$, and HD$_{99}$ to cross-sectional FreeSurfer surfaces, the silver standard in the field~\citep{cruzDeepCSR3DDeep2020,maCortexODELearningCortical2022,lebratCorticalFlowDiffeomorphicMesh,hoopesTopoFitRapidReconstruction}. These results are based on our ADNI and OASIS test sets, separated by WM and pial surfaces.} All methods achieve excellent accuracy with an average HD$_{90}$ mostly below 0.5mm, i.e., half the image resolution, and an ASSD of 0.2mm and below. Although the margins are comparably small, V2C-Long still yields the highest accuracy among the deep learning-based methods in all metrics on both datasets, except for the 99-percentile Hausdorff distance, where CF$^{++}$ is superior. 
{\revtwo Notably, the spherical registration reduces the accuracy of V2C-Flow surfaces, e.g., by 0.017mm (ASSD) for WM surfaces on ADNI.}
We also include FS-Long in \Cref{tab:adni_reconstruction} for completeness. However, FS-Long intrinsically uses FreeSurfer methods {\rev (not just for training but also to process test scans); hence, the comparison to learning-based methods is not entirely fair}. 

For a visual comparison of V2C-Long and FS-Long, we show cortical reconstructions from two visits of a subject in the ADNI test set in  \Cref{fig:qualitative} {\rev (see Supplementary Figure~3 for qualitative results from all evaluated methods)}.
The highlighted regions in \Cref{fig:qualitative} show anatomically implausible reconstructions by FS-Long in both visits and the within-subject template.  
In contrast, V2C-Long circumvents the fissures in the within-subject template and yields visually smooth and consistent surfaces at both visits.

{\rev
To provide an additional, FreeSurfer-independent assessment of the reconstruction, we evaluate the accuracy of V2C-Long, V2CC, and FS-Long to differentiate between diagnostic groups, i.e., Alzheimer's disease and cognitively normal, in our ADNI test set based on derived cortical thickness measures. This analysis is grounded in the known association of Alzheimer's disease with cortical atrophy~\citep{schwarz2016alzheimers}. From the cortical thickness obtained from the respective method, we computed vertex-wise Z-scores~\citep{Tahedl2020zscores}, indicating the deviation from the norm in an age-matched, cognitively normal reference cohort (stratified by 10-year age brackets). For the estimation of the mean and standard deviation in each age cohort, we considered only initial scans to avoid bias toward individuals with many scans. Using the average Z-score across the cortex (excluding the unknown medial region), we calculated the Area Under the Curve (AUC) to classify the two groups. V2C-Long achieves the highest diagnostic accuracy from cortical thickness measures with an AUC of 0.83, followed by V2CC with an AUC of 0.81 and FS-Long with an AUC of 0.77.
}

\subsection{Ablation study} \label{sec:ablation}

{\rev
In \Cref{tab:ablation}, we validate the design choices made in V2C-Long. In particular, we compare the realization of each of V2C-Long's building blocks, i.e., the within-subject template creation and the within-subject template deformation, to alternative approaches. For the within-subject template creation, we consider approaches that are based on the FsAverage template as alternatives to V2C-Flow, i.e., V2CC and TopoFit, as well as FreeSurfer's base images (FS-Base). For the within-subject template deformation, we ablate the V2C-Flow model and replace it with CorticalFlow (CF) deformation blocks. As this requires substantial modification of the CF$^{++}$ method, we refer to it as ``CF blocks'' instead of ``CF$^{++}$'' in \Cref{tab:ablation}. We did not employ V2CC and TopoFit as the within-subject template-deformation model as their training relies on cross-sectional ground-truth correspondences, which cannot be easily adapted for the longitudinal template deformation step.

\begin{table}[t]
\setlength{\tabcolsep}{2.5pt}
\renewcommand{\arraystretch}{1.1}
\renewcommand\bfdefault{b}% rather than bx
\small
\centering
\caption{\rev Evaluation of different realizations of the within-subject template creation, aggregation, and deformation, as well as the vertex feature (VF)-enhancement. We report mean{\scriptsize$\pm$SD} values over all subjects (MCVar, CThVar) respectively all scans (SIF, HD$_{90}$) in the ADNI test set. Best values are \textbf{highlighted}.}
\begin{adjustbox}{center}
\begin{tabular}{@{}lllcccccccc@{}}
\toprule
\multirow[b]{2}{*}[+4pt]{\makecell[l]{Within-sub.\\template\\creation}} &
\multirow[b]{2}{*}[+4pt]{\makecell[l]{Within-sub.\\template\\aggregation}} & 
\multirow[b]{2}{*}[+4pt]{\makecell[l]{\rev Within-sub.\\ \rev template\\ \rev deformation}}& 
\multirow[b]{2}{*}[0pt]{VF} & 

\multicolumn{2}{c}{WM surface} &  
\multicolumn{2}{c}{Pial surface} &  
\multicolumn{2}{c}{Both}\\

\cmidrule(lr){5-6}
\cmidrule(lr){5-6}
\cmidrule(lr){7-8}
\cmidrule(lr){7-8}
\cmidrule(lr){9-10}
\cmidrule(lr){9-10}
 & 
 & 
 &
 &
MCVar$\downarrow$ & 
\%SIF$\downarrow$ & 
MCVar$\downarrow$ & 
\%SIF$\downarrow$ & 
CThVar$\downarrow$ &
$\textsf{HD}_{90}$$\downarrow$ 
\\
\midrule

FS-Base & Median image & \rev V2C-Flow & $\times$ &
.021\scriptsize\ensuremath{\pm}.008 & 
1.062\scriptsize\ensuremath{\pm}0.334 & 
.013\scriptsize\ensuremath{\pm}.005 & 
3.070\scriptsize\ensuremath{\pm}1.035 & 
.017\scriptsize\ensuremath{\pm}.010 &
.398\scriptsize\ensuremath{\pm}.719 &
\\

V2C-Flow & Median mesh & \rev V2C-Flow & $\times$ &
.035\scriptsize\ensuremath{\pm}.013 &
1.330\scriptsize\ensuremath{\pm}0.867 &
.022\scriptsize\ensuremath{\pm}.008 &
2.810\scriptsize\ensuremath{\pm}1.121 &
.019\scriptsize\ensuremath{\pm}.009 &
.411\scriptsize\ensuremath{\pm}.313 &
\\

\rev TopoFit & \rev Mean mesh & \rev V2C-Flow & $\times$ &
\rev .027\scriptsize\ensuremath{\pm}.011 &
\rev \textbf{0.260}\scriptsize\ensuremath{\pm}1.666 &
\rev .022\scriptsize\ensuremath{\pm}.010 &
\rev 2.324\scriptsize\ensuremath{\pm}1.824 &
\rev .020\scriptsize\ensuremath{\pm}.009 &
\rev .447\scriptsize\ensuremath{\pm}.255 &
\\

V2CC & Mean mesh & \rev V2C-Flow  & $\times$ &
.024\scriptsize\ensuremath{\pm}.009 & 
0.344\scriptsize\ensuremath{\pm}0.412 & 
.016\scriptsize\ensuremath{\pm}.006 & 
2.526\scriptsize\ensuremath{\pm}1.100 & 
.019\scriptsize\ensuremath{\pm}.008 &
\textbf{.392}\scriptsize\ensuremath{\pm}.078 & 
\\

V2C-Flow & Mean mesh & \rev V2C-Flow  & $\times$ &
.022\scriptsize\ensuremath{\pm}.027 & 
0.649\scriptsize\ensuremath{\pm}0.439 & 
.013\scriptsize\ensuremath{\pm}.020 & 
2.323\scriptsize\ensuremath{\pm}0.926 & 
.018\scriptsize\ensuremath{\pm}.012 &
.401\scriptsize\ensuremath{\pm}.623 & 
\\

\rev V2C-Flow & 
\rev Mean mesh & 
\rev CF blocks & 
\rev $\times$ &
\rev .036\scriptsize\ensuremath{\pm}.013 &
\rev  0.659\scriptsize\ensuremath{\pm}0.275 &
\rev .024\scriptsize\ensuremath{\pm}.010 &
\rev \textbf{2.076}\scriptsize\ensuremath{\pm}0.823 &
\rev .020\scriptsize\ensuremath{\pm}.010 &
\rev .478\scriptsize\ensuremath{\pm}.258 &
\\

V2C-Flow & Mean mesh & \rev V2C-Flow  & $\checkmark$ &
\textbf{.020}\scriptsize\ensuremath{\pm}.020 & 
0.643\scriptsize\ensuremath{\pm}0.375 & 
\textbf{.012}\scriptsize\ensuremath{\pm}.015 & 
2.151\scriptsize\ensuremath{\pm}0.921 & 
\textbf{.016}\scriptsize\ensuremath{\pm}.010 &
.410\scriptsize\ensuremath{\pm}.743 & 
\\
\bottomrule
\end{tabular}

\end{adjustbox}
\\

\label{tab:ablation}
\end{table}

\noindent
\textbf{Within-subject template creation and aggregation.}
Compared to FreeSurfer's base images (FS-Base), we found the mean aggregation of the V2C-Flow meshes to reduce the number of self-intersecting faces in the output surfaces. Similarly, mimicking FreeSurfer's median aggregation in mesh space was detrimental in our ablation experiments as it considerably increased MCVar and SIF measures. In terms of run time, creating FreeSurfer's base template (\texttt{recon-all~-base}) takes around 4 hours on an Intel i7 CPU running at 3.60GHz for five visits. In V2C-Long, we obtain the within-subject template from the same number of visits within less than 10 seconds using a recent Nvidia A100 GPU,
requiring around 8.5GB of VRAM at inference time. If no GPU is available, V2C-Long can still be run on the CPU. This takes around one minute per visit on an Intel i7 at 3.60GHz. Replacing the V2C-Flow template-creation model with V2CC improved the HD$_{90}$ reconstruction accuracy by around 0.01mm at the cost of slightly less consistent surfaces in our experiments.
The TopoFit within-subject template-creation model yielded the lowest number of self-intersecting faces in WM surfaces, but it did not improve over the V2C-Flow model in the remaining metrics.
Unfortunately, the current limitations of GPU VRAM impede the use of vertex features from all four/seven graph deformation blocks in V2CC/TopoFit.

\noindent
\textbf{Within-subject template deformation.}
Replacing the V2C-Flow deformation model in the second step of V2C-Long with comparable CF blocks led to a slight reduction of the SIF score on the pial surfaces; however, it resulted in worse outcomes across all other metrics in \Cref{tab:ablation}. Finally, we found the enhancement of within-subject templates with vertex features (VF) in V2C-Long to have a slightly positive impact on the consistency of WM and pial surfaces and cortical thickness measures. 
}

\subsection{Longitudinal cortical thickness in Alzheimer's disease} \label{sec:groupstudy-long}

\begin{figure}[t]
    \centering
    \includegraphics[width=\textwidth]{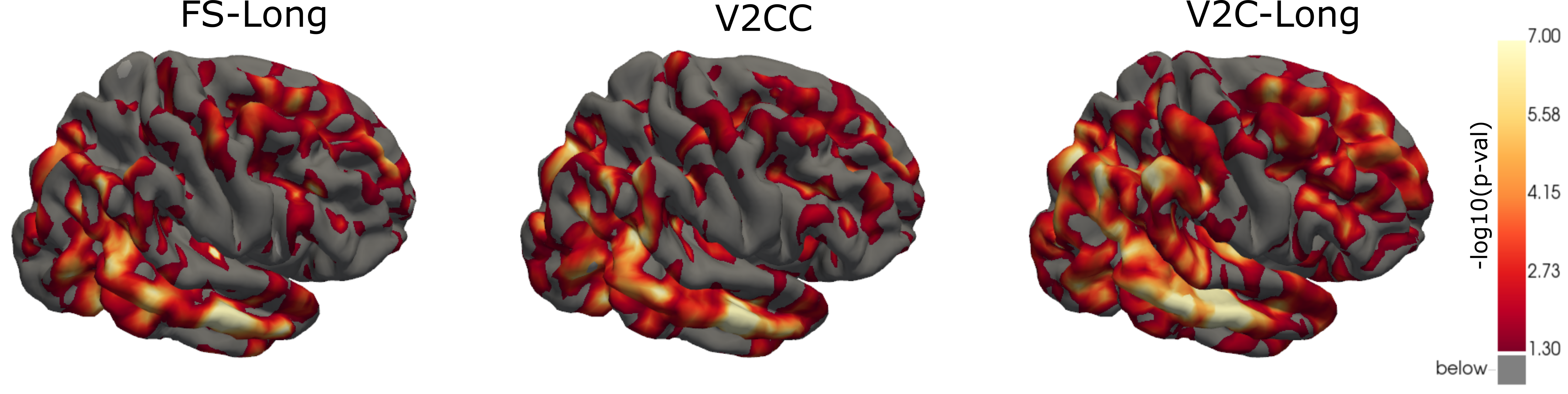}
    \caption{Differences in longitudinal cortical thickness between stable AD subjects ($n=71$) and healthy controls ($n=193$). We show uncorrected negative log10(p-value)-maps 
    %($-\text{log10}(0.05)\approx 1.3$) 
    on FsAverage --- based on cortical surfaces from FS-Long (v7.2), V2CC, and V2C-Long.}
    \label{fig:groupstudy}
\end{figure}

In \Cref{fig:groupstudy} (see Supplementary Figure~1 for the {\rev medial view}), we compare the pattern of longitudinal cortical thickness in patients affected by Alzheimer's disease (AD) with a cognitively normal control group. 
To this end, we select $n=71$ AD subjects with a stable diagnosis and $n=193$ healthy controls from our ADNI test set and perform a mass-univariate analysis of longitudinal cortical thickness (CTh) measurements with the vertex-wise LME regression model described in \Cref{sec:mle}. We plot uncorrected two-tailed p-values of the t-statistics of the regression effect $\beta_3$, i.e., the diagnosis. 
{
\rev
We compare the atrophy patterns among the three methods with the best longitudinal consistency (cf.~\Cref{tab:adni_correspondence}), i.e., FS-Long, V2CC, and V2C-Long. All three methods highlight similar regions in the temporal, parietal, and frontal lobes, as well as the precuneus and the posterior cingulate cortex.
Exceptions are the isthmus cingulate and parts of the orbitofrontal cortex, where V2C-Long does not indicate group differences based on $p<0.05$ in contrast to the other two methods. V2C-Long, on the other hand, detects generally larger areas in the frontal, temporal, and parietal lobes, i.e., lower p-values, providing stronger evidence for group differences.
}

\section{Discussion}

This work introduced V2C-Long for longitudinal cortical surface reconstruction with spatiotemporal correspondence of surface points. 
The developed model yielded highly consistent results across visits and high accuracy across datasets, indicating reliable and robust reconstruction performance.

\textbf{Longitudinal reconstruction of cortical surfaces.}
Achieving consistent cortex reconstruction from MRI, where individual vertices are placed at precise anatomical locations, is challenging due to the {\revtwo complex folding patterns}, the variability across subjects, and the limiting image resolution.
We demonstrated that V2C-Long exhibits a compelling longitudinal consistency (cf. \Cref{sec:rec-consistency}) as determined by quantitative scores and qualitative inspection of reconstructed surfaces on internal (ADNI) and external (OASIS) study data. {\revtwo The highest region-wise} inconsistencies occurred for all methods in small and thin regions in the cingulate and insular cortex (cf.~\Cref{fig:f1parc}), which is understandable as inconsistencies at the parcellation boundaries have a larger impact in these cases relative to the region size. {\rev On the vertex level, we observed considerable inconsistencies in longitudinal mean curvature in CF$^{++}$ and FS-Long in the occipital lobe, especially in pial surfaces (cf.~\Cref{fig:mcvar}). V2C-Long, on the other hand, {\revtwo exhibited} very low longitudinal mean curvature variance (MCVar) across the entire cortex, with the largest inconsistencies again {\revtwo occurring} around the cingulate cortex and the insula.} The evaluation of test-retest reliability (cf. \Cref{sec:trt}), which can be seen as short-term longitudinal consistency, led to similar results as the evaluation on ADNI and OASIS. In this regard, V2C-Long also {\revtwo improved} over the best available reconstruction methods. At the same time, V2C-Long does not sacrifice accuracy; the model is at the forefront of all implemented deep learning-based architectures (cf.~\Cref{sec:rec-accuracy}). {\rev The performance metrics of all neural networks on OASIS, of which no data  has been used during training, were principally consistent with ADNI. This indicates good generalization of trained models across studies.} 
{\rev For comparison}, we implemented recent template-based cortex reconstruction methods with different neural network architectures {\rev (CNN: CF$^{++}$; CNN \& GNN: TopoFit, V2CC, V2C-Flow) and loss functions (L1: V2CC; Chamfer: CF$^{++}$; restricted Chamfer: TopoFit; curvature-weighted Chamfer: V2C-Flow)}. These methods, however, were designed for cross-sectional data 
and our experiments revealed that their longitudinal consistency is not competitive with dedicated longitudinal methods such as FS-Long. The only cross-sectional method {\revtwo approaching the performance of} FS-Long is V2CC, which has been trained in a supervised manner to create correspondences based on previously registered and re-sampled FreeSurfer surfaces. {\rev Nonetheless, V2C-Long {\revtwo outperformed} FS-Long and V2CC without requiring any surface-based pre- or post-processing. In addition, the learning-based approach makes V2C-Long more versatile than FS-Long. The model could be re-trained or fine-tuned for particular scenarios, e.g., higher magnetic field strengths above 3T}.

\textbf{Creation of within-subject templates.}
From comparing V2C-Long with its direct architectural baseline, V2C-Flow, we deduce that creating within-subject templates is crucial for first-rate longitudinal consistency. Albeit within-subject templates are also used by FS-Long to initialize the reconstruction at the individual visits, V2C-Long computes the within-subject templates directly in mesh space. Eventually, this leads to a speed-up in the within-subject template creation from hours to seconds compared to the group-wise registration in FS-Long (cf.~\Cref{sec:ablation}). 
{\rev In \Cref{fig:qualitative}, we observed that the mesh-based aggregation prevents implausible anatomical errors in the shown case, i.e., fissures in the pial cortical surface}. These artifacts likely result from converting a defective segmentation into surface meshes, which is circumvented entirely in V2C-Long since the sphere-like topology is already engraved into the input template. {\rev Supported by \Cref{fig:qualitative}, Supplementary Figure~3, and the work by \cite{lebrat_robust_isbi_2023}, we have good evidence that the geometric template-deformation in V2C-Long is robust and yields plausible results even for challenging input images. V2C-Long benefits from the template deformation not only in the ultimate reconstruction but also during the within-subject template creation.}

\textbf{Architectural design choices and training setup.}
Compared to the improvement over existing methods, we found architectural design choices to have a minor impact on the reconstruction quality and consistency (cf.~\Cref{sec:ablation}). Yet, the best results were obtained by enhancing the within-subject template with deep vertex features and mean mesh aggregation. The GNN allows us to process information related to individual vertices and augment the input templates beyond bare vertex coordinates, which is impossible in a purely CNN-based approach. Aggregating the meshes via a median operation similar to FreeSurfer's median image~\citep{reuterWithinsubjectTemplateEstimation2012} or employing FS-Base as a starting point for the longitudinal reconstruction is technically feasible but not recommended as we observed a dramatic increase in self-intersections in reconstructed WM surfaces. Our implementation allows for joint training and prediction of all four cortical surfaces simultaneously on a single GPU (cf.~Supplementary Table~1). In contrast, TopoFit and CF$^{++}$ train a single model for each cortical surface, complicating their practical usability. In addition, the joint reconstruction permits the incorporation of anatomical constraints between WM and pial surfaces, which proved beneficial in avoiding their intersection~\citep{Bongratz2024v2cflow}.

\noindent
\textbf{Longitudinal group analyses.}
The association of Alzheimer's disease and cortical thickness has been well studied in the literature~\citep{Singh2006patternsofcorticalthinning,Du2006,Risacher2010,schwarz2016alzheimers}. To conduct these studies, extracting cortical thickness measures with FreeSurfer is a prominent, if not the most prevalent, approach. Our experiments replicated these results for the first time in a longitudinal setting using deep-learning methods (cf.~\Cref{sec:groupstudy-long}). 
{\rev 
Overall, the regions highlighted by FS-Long, V2CC, and V2C-Long (cf.~\Cref{fig:groupstudy} and Supplementary Figure~1) based on $p<0.05$ are consistent with existing research on cortical changes in Alzheimer's disease, which reported a broad pattern of cortical atrophy most significant in the temporal lobe, the temporoparietal junction, the posterior cingulate, and the precuneus~\citep{Du2006}. Nevertheless, we found slight differences between the three methods. V2CC and FS-Long highlighted group differences in the isthmus of the cingulate gyrus and the orbitofrontal cortex, which V2C-Long {\revtwo did} not feature. Instead, V2C-Long {\revtwo carved} out more significant evidence for atrophy in large parts of the frontal, temporal, and parietal lobes than the other two methods. 
}
{\rev 
The better diagnostic accuracy in V2C-Long (AUC 0.83) compared to V2CC (AUC 0.81) and FS-Long (AUC 0.77), cf.~\Cref{sec:rec-accuracy}, further underscores the capability of V2C-Long to distinguish between AD patients and cognitively normal controls based on estimated cortical thickness. 
In contrast to FS-Long and V2CC, V2C-Long requires no spherical inflation and re-sampling for such vertex-based longitudinal group analyses (required as a post-processing step in FS-Long and as pre-processing in V2CC). 
Instead, we conducted all analyses with the raw output surfaces of V2C-Long, which reduces potential sources of error and computation time to a minimum. 
}

{\revtwo
\textbf{Correspondences and registration.}
With V2C-Long, we obtain spatiotemporal vertex correspondences across subjects (cross-sectional) and visits (longitudinal) that match anatomical locations on the FsAverage template. The cross-sectional correspondences emerge from the shape alignment with the curvature-weighted Chamfer loss, cf. Supplementary Figures~5 and~6, which we found to be sufficient for the experiments conducted in this paper. However, such unsupervised alignment is not equivalent to an explicit curvature-based registration. For specific applications, such as atlas-based parcellation, V2C-Long might still benefit from a subsequent registration step. Based on V2C-Long's inherent correspondence, the template icosphere can directly be used for this purpose, without the need for spherical inflation, similar to the parcellation in V2C-Flow~\citep{Bongratz2024v2cflow}. Nevertheless, we found that post hoc registration tends to reduce the accuracy of reconstructed surfaces (see \Cref{tab:adni_reconstruction}), likely due to the interpolation of vertex coordinates. Additionally, we demonstrated in \Cref{tab:adni_correspondence} that V2C-Long's longitudinal correspondences, which were the focus of this work, are significantly superior to those obtained through traditional surface registration. 
}

\noindent
\textbf{Limitations.}
As with all supervised learning methods, V2C-Long relies on the availability and correctness of ground-truth datasets. Compared to most other applications, however, obtaining manual cortical surface meshes by human experts is infeasible. As it has become a standard in the field, we {\revtwo used} cross-sectional FreeSurfer as a reference for model training and evaluation of reconstruction accuracy~\citep{cruzDeepCSR3DDeep2020,lebratCorticalFlowDiffeomorphicMesh,maCortexODELearningCortical2022}. Although we tried to remove scans with severe processing artifacts, we cannot preclude that errors made by FreeSurfer affect our models and confound the evaluation presented in this paper. The consistency assessment, however, requires no reference standard and provides an additional perspective usually not reported in prior studies.
Our training set comprises more than 3.7K T1w MRIs from cognitively normal subjects as well as subjects diagnosed with Alzheimer's disease and mild cognitive impairment, making it the largest training set in the literature for cortex reconstruction models. Still, future work should investigate the performance under other neurodegenerative conditions or brain tumors. Self-intersections could potentially also confound downstream applications. We did not find self-intersections in V2C-Long obstructive, as we achieved excellent results in benchmark experiments and downstream applications with the raw output. 
To guarantee flawless compatibility with existing neuroimaging tools, however, we will work towards reducing the number of self-intersecting faces in V2C-Long, especially on the pial surface. Nevertheless, we believe that our model can already be used for a broad range of applications \emph{as is}, and particular failure cases can be addressed via re-training or fine-tuning.

\noindent
\textbf{Conclusion.}
In conclusion, we introduced V2C-Long, the first dedicated deep-learning method for longitudinal cortex reconstruction and within-subject template creation. V2C-Long uses deep deformation fields to establish a strong inherent spatiotemporal correspondence between cortical surfaces, rendering them directly comparable without post-processing. {\rev Our experiments on two large longitudinal brain MRI studies validated the accuracy and consistency, demonstrating a substantial improvement over previous methods. 
We provided stronger evidence of longitudinal cortical atrophy in Alzheimer's disease and higher diagnostic accuracy than FreeSurfer.
Our results show the potential for V2C-Long to enhance future longitudinal neuroimage analyses, and the developed model offers researchers a valuable tool to find more subtle associations with brain structure.}

\section*{Data and Code Availability}
This study uses public data from the Alzheimer's Disease Neuroimaging Initiative (ADNI, \url{http://adni.loni.usc.edu}) and from the Open Access Series of Imaging Studies (OASIS-3: Longitudinal Multimodal Neuroimaging, \url{https://sites.wustl.edu/oasisbrains}). Access is obtained through the online application forms provided at the linked URLs. Our code is available on GitHub with the link mentioned in the manuscript.

\section*{Author Contributions}
Fabian Bongratz: Conceptualization, methodology, software, validation, formal analysis, data curation, writing --- original draft, writing —-- review \& editing, and visualization. 
Jan Fecht:  Methodology, software, validation, formal analysis, investigation, writing—original draft, and visualization.
Anne-Marie Rickmann: Conceptualization, software, data curation, and writing --— review \& editing.
Christian Wachinger: Conceptualization, resources, writing --— review \& editing, supervision, project administration, and funding acquisition.

\section*{Ethics Statement}
This work uses data from the Alzheimer's disease neuroimaging initiative and (ADNI, \url{http://adni.loni.usc.edu}) and from the Open Access Series of Imaging Studies (OASIS-3: Longitudinal Multimodal Neuroimaging, \url{https://sites.wustl.edu/oasisbrains}).
Participants in both studies gave written informed consent in accordance with the study's ethical guidelines. Data acquisition in ADNI adheres to the Declaration
of Helsinki; the study protocol is available at \url{https://adni.loni.usc.edu/wp-content/themes/freshnews-dev-v2/documents/clinical/ADNI-1_Protocol.pdf}. OASIS-3 was approved by Washington University’s Institutional Review Board~\citep{lamontagneOASIS3LongitudinalNeuroimaging2019}. Participants in the test-retest database (TRT, \url{https://doi.org/10.6084/m9.figshare.929651}) gave written informed consent following the guidelines of the Stanford University Institutional Review Board~\citep{maclarenReliabilityBrainVolume2014}. 

\section*{Declaration of Competing Interests}

The authors declare that they have no conflict of interest.

\section*{Acknowledgements}
This research was partially supported by the German Research
Foundation. We gratefully acknowledge the computational resources provided by the
Leibniz Supercomputing Centre (www.lrz.de). 
The authors would also like to thank Lennart Bastian for initial fruitful discussions.
Data collection and sharing for the Alzheimer's Disease Neuroimaging Initiative (ADNI) is
funded by the National Institute on Aging (National Institutes of Health Grant U19
AG024904). The grantee organization is the Northern California Institute for Research and
Education.

\printbibliography

\clearpage

\appendix

\section{Appendix} \label{sec:appendix}

We prove \Cref{thm:template} by induction. Assume we have input scans $X_{\rev 0,1,\ldots,K_i}$ from a certain subject $i$ with $K_i+1$ visits. Then, the initial value problem
\begin{equation*}
    \frac{d\mathcal{T}_i(t)}{dt} = \bar{f}(t, X_{\rev 0,1,\ldots,K_i}, \mathcal{T}_i(t)); \quad \mathcal{T}_i(0) = \mathcal{T}
\end{equation*}
describes the deformation of the generic input template $\mathcal{T}$ under the aggregated mean flow field $\bar{f}$ across all visits. Further, we have
\begin{equation*}
    \mathcal{T}_i(0) = \frac{1}{K_i + 1}\sum_{j=0}^{K_i} \mathcal{V}_{i,j}(0)
\end{equation*}
since $\mathcal{V}_{i,j}(0) = \mathcal{T}$ $\forall i,j$ in V2C-Flow, which proves our statement for the simple case $t=0$.
For each next forward Euler integration step $s+1 \geq 1$, $s\in \mathbb{N}^0$, with the induction hypothesis $\mathcal{T}_{i}(t) = \frac{1}{K_i+1}\sum_{j=0}^{K_i}\mathcal{V}_{i,j}(t)$, we obtain for $t=hs\geq0$:
\begin{align*}
    \mathcal{T}_{i,s+1} &= \mathcal{T}_{i,s} + h\bar{f}(hs, X_{\rev 0,1,\ldots,K_i}, \mathcal{T}_{i,s}) \\
    &= \frac{1}{K_i + 1} \sum_{j=0}^{K_i} \mathcal{V}_{i,j,s} + \frac{h}{K_i + 1}\sum_{j=0}^{K_i}{f}(hs, X_j, \mathcal{V}_{i,j,s}) \\
    %&= \frac{1}{K_i + 1} \sum_{j=0}^{K_i} \mathcal{V}_{i,j,s} + hf(hs, X_j, \mathcal{V}_{i,j,s}) \\
    &= \frac{1}{K_i + 1} \sum_{j=0}^{K_i} \mathcal{V}_{i,j,s+1} \\
    \qed
\end{align*}

\clearpage

\section{Supplementary Material}
\renewcommand{\figurename}{Supplementary Figure}
\setcounter{figure}{0}
\renewcommand{\tablename}{Supplementary Table}
\setcounter{table}{0}

\begin{figure}[htb]
    \centering
    \includegraphics[width=\textwidth]{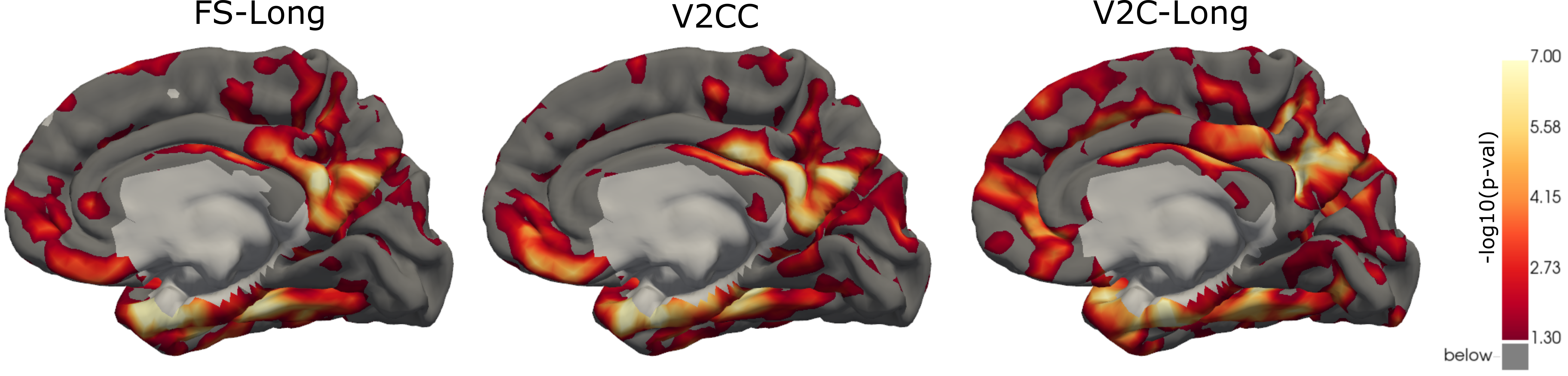}
    \caption{{\rev Medial view} on differences in longitudinal cortical thickness between stable AD subjects ($n=71$) and healthy controls ($n=193$). We show uncorrected negative log10(p-value)-maps 
    on FsAverage --- based on meshes from FS-Long (v7.2), V2CC, and V2C-Long.}
    \label{fig:group-sagittal}
\end{figure}

\begin{figure}[htb]
    \centering
    \includegraphics[width=\textwidth]{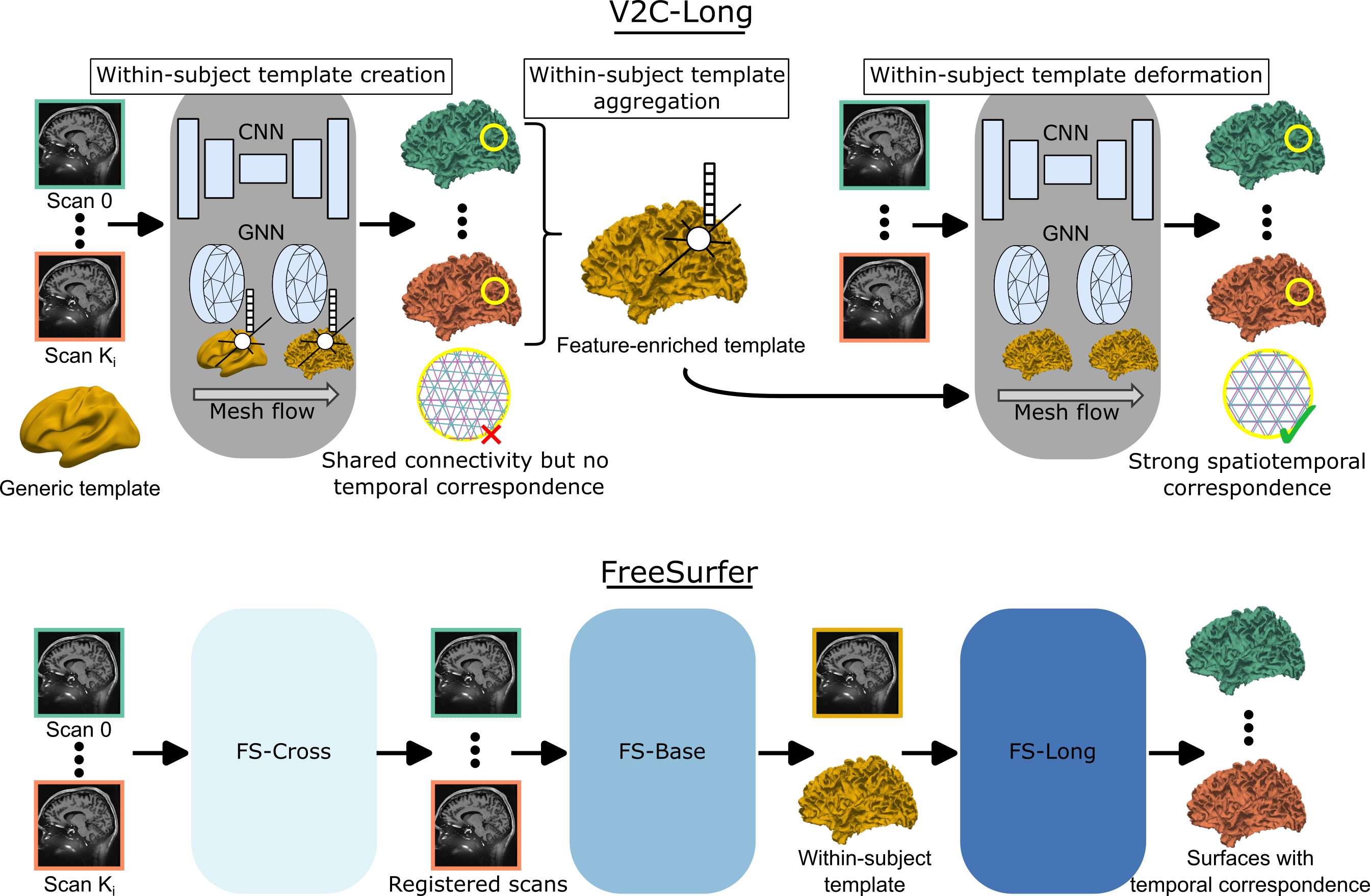}
    \caption{\rev Conceptual comparison of V2C-Long with the longitudinal FreeSurfer pipeline.}
    \label{fig:architecture-extended}
\end{figure}

\begin{table}[h]

\centering
\caption{Results from a joint V2C-Long model trained for the left (lh) and right (rh) hemisphere. We report accuracy (ASSD) and consistency (ParcF1) metrics by surface and hemisphere. Values are mean{\scriptsize$\pm$SD} over all subjects (ParcF1) respectively all scans (ASSD) in our ADNI test set.}
\label{tab:joint-model}
\begin{tabular}{@{}cccccccc@{}}
\multicolumn{2}{c}{WM surface lh} &  \multicolumn{2}{c}{Pial surface lh} & \multicolumn{2}{c}{WM surface rh} &  \multicolumn{2}{c}{Pial surface rh} \\
 \cmidrule(lr){1-2}
 \cmidrule(lr){1-2}
 \cmidrule(lr){3-4}
 \cmidrule(lr){3-4}
 \cmidrule(lr){5-6}
 \cmidrule(lr){5-6}
 \cmidrule(lr){7-8}
 \cmidrule(lr){7-8} 
  ASSD$\downarrow$ & ParcF1$\uparrow$ & ASSD$\downarrow$ & ParcF1$\uparrow$ & ASSD$\downarrow$ & ParcF1$\uparrow$ & ASSD$\downarrow$ & ParcF1$\uparrow$ \\
 \toprule
 0.170{\scriptsize$\pm$0.055} & 0.967{\scriptsize$\pm$0.013} & 0.163{\scriptsize$\pm$0.059} &  0.962{\scriptsize$\pm$0.015} & 0.169{\scriptsize$\pm$ 0.053} & 0.969{\scriptsize$\pm$0.011} & 0.162{\scriptsize$\pm$0.060} & 0.963{\scriptsize$\pm$0.014}\\
 
 \bottomrule
\end{tabular}

\end{table}

\LTcapwidth=\textwidth
\begin{longtable}{l}
\caption{Destrieux atlas regions in counter-clockwise order.}
\label{tab:destrieux-regions}
\endfirsthead
\caption{Destrieux atlas regions in counter-clockwise order (continued).} 
\endhead
   
        G\_insular\_short\\ 
        G\_occipital\_middle\\ 
        G\_occipital\_sup\\ 
        G\_oc-temp\_lat-fusifor\\ 
        G\_oc-temp\_med-Lingual\\ 
        G\_oc-temp\_med-Parahip\\ 
        G\_orbital\\ 
        G\_pariet\_inf-Angular\\ 
        G\_pariet\_inf-Supramar\\ 
        G\_parietal\_sup\\ 
        G\_postcentral\\ 
        G\_precentral\\ 
        G\_precuneus\\ 
        G\_rectus\\ 
        G\_subcallosal\\ 
        G\_temp\_sup-G\_T\_transv\\ 
        G\_temp\_sup-Lateral\\ 
        G\_temp\_sup-Plan\_polar\\ 
        G\_temp\_sup-Plan\_tempo\\ 
        G\_temporal\_inf\\ 
        G\_temporal\_middle\\ 
        Lat\_Fis-ant-Horizont\\ 
        Lat\_Fis-ant-Vertical\\ 
        Lat\_Fis-post\\ 
        Pole\_occipital\\ 
        Pole\_temporal\\ 
        S\_calcarine\\ 
        S\_central\\ 
        S\_cingul-Marginalis\\ 
        S\_circular\_insula\_ant\\ 
        S\_circular\_insula\_inf\\ 
        S\_circular\_insula\_sup\\ 
        S\_collat\_transv\_ant\\ 
        S\_collat\_transv\_post\\ 
        S\_front\_inf\\ 
        S\_front\_middle\\ 
        S\_front\_sup\\ 
        S\_interm\_prim-Jensen\\ 
        S\_intrapariet\_and\_P\_trans\\ 
        S\_oc\_middle\_and\_Lunatus\\ 
        S\_oc\_sup\_and\_transversal\\ 
        S\_occipital\_ant\\ 
        S\_oc-temp\_lat\\ 
        S\_oc-temp\_med\_and\_Lingual\\ 
        S\_orbital\_lateral\\ 
        S\_orbital\_med-olfact\\ 
        S\_orbital-H\_Shaped\\ 
        S\_parieto\_occipital\\ 
        S\_pericallosal\\ 
        S\_postcentral\\ 
        S\_precentral-inf-part\\ 
        S\_precentral-sup-part\\ 
        S\_suborbital\\ 
        S\_subparietal\\ 
        S\_temporal\_inf\\ 
        S\_temporal\_sup\\ 
        S\_temporal\_transverse\\ 
        Unknown\\ 
        G\_and\_S\_frontomargin\\ 
        G\_and\_S\_occipital\_inf\\ 
        G\_and\_S\_paracentral\\ 
        G\_and\_S\_subcentral\\ 
        G\_and\_S\_transv\_frontopol\\ 
        G\_and\_S\_cingul-Ant\\ 
        G\_and\_S\_cingul-Mid-Ant\\ 
        G\_and\_S\_cingul-Mid-Post\\ 
        G\_cingul-Post-dorsal\\ 
        G\_cingul-Post-ventral\\ 
        G\_cuneus\\ 
        G\_front\_inf-Opercular\\ 
        G\_front\_inf-Orbital\\ 
        G\_front\_inf-Triangul\\ 
        G\_front\_middle\\ 
        G\_front\_sup\\ 
        G\_Ins\_lg\_and\_S\_cent\_ins
    
\end{longtable}

\begin{figure}
    \centering
    \includegraphics[width=0.8\textwidth]{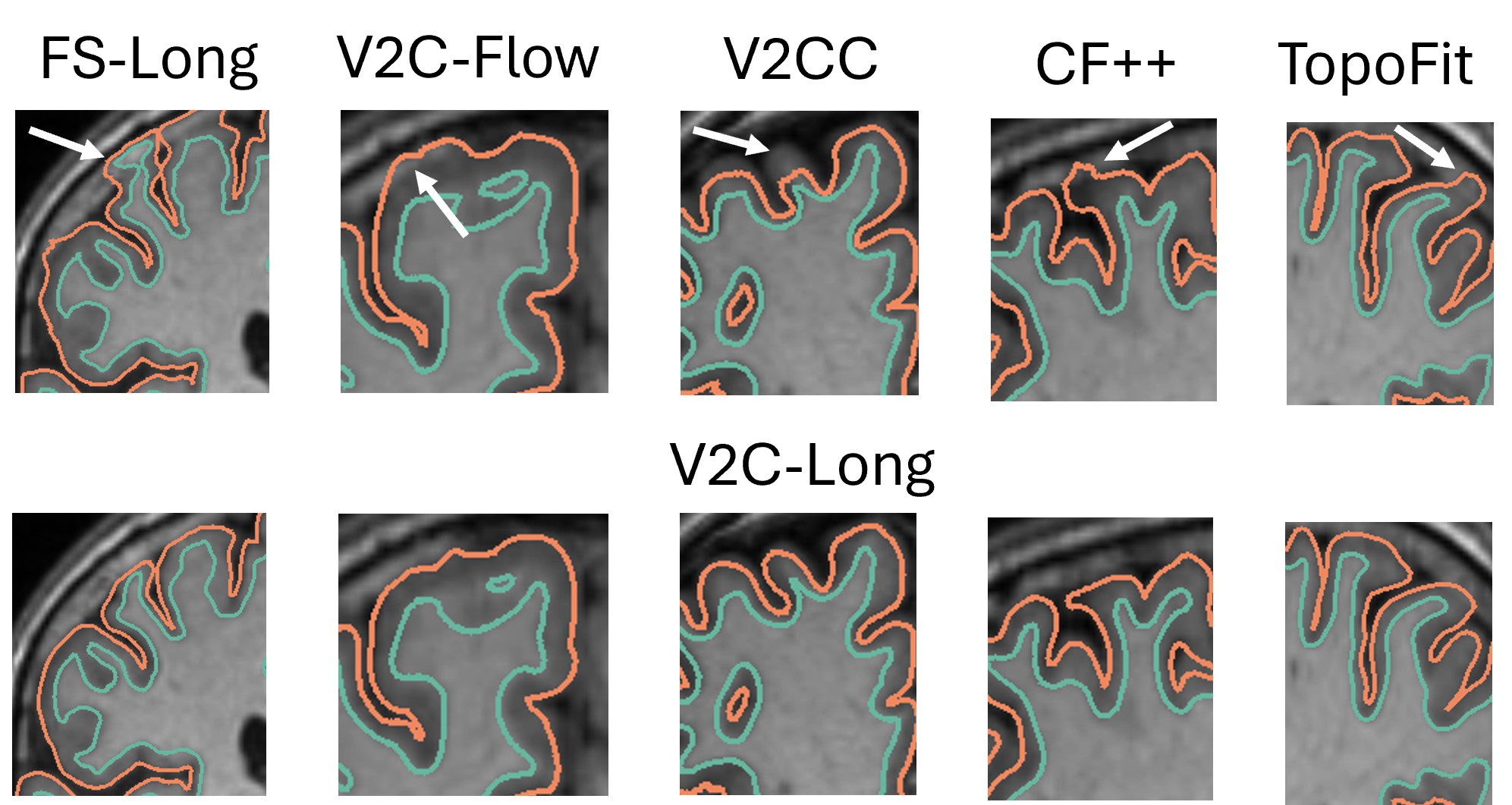}
    \caption{\rev Failure cases of existing methods (indicated by white arrows, top row) and corresponding V2C-Long segmentation (bottom row).}
    \label{fig:failure}
\end{figure}

\begin{figure}[htbp]
    \centering
    \includegraphics[width=0.8\textwidth]{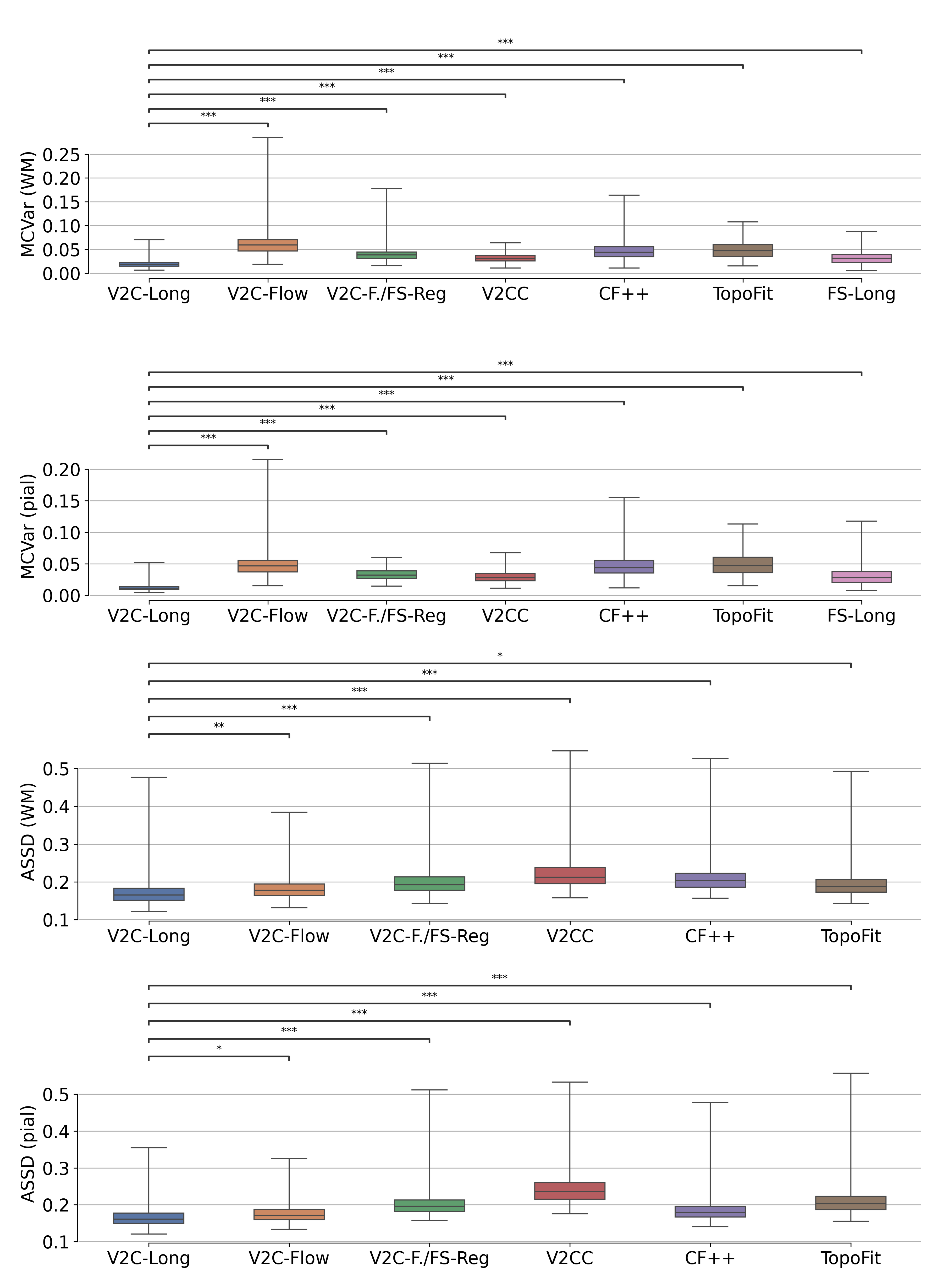}
    \caption{\rev Longitudinal consistency (MCVar) and reconstruction accuracy (ASSD) for white matter (WM) and pial surfaces in our ADNI test set. Boxplots show the median as the central line, boxes representing the quartiles, and whiskers extending to 10 times the interquartile range. Stars indicate significance levels based on paired t-tests between the scores of V2C-Long and the other methods. $^{***}: p<\num{1e-10}$, $^{**}: p<\num{1e-5}$, $^{*}: p<0.01$.}
    \label{fig:results-plots}
\end{figure}

\begin{figure}
    \centering
    \includegraphics[width=\textwidth]{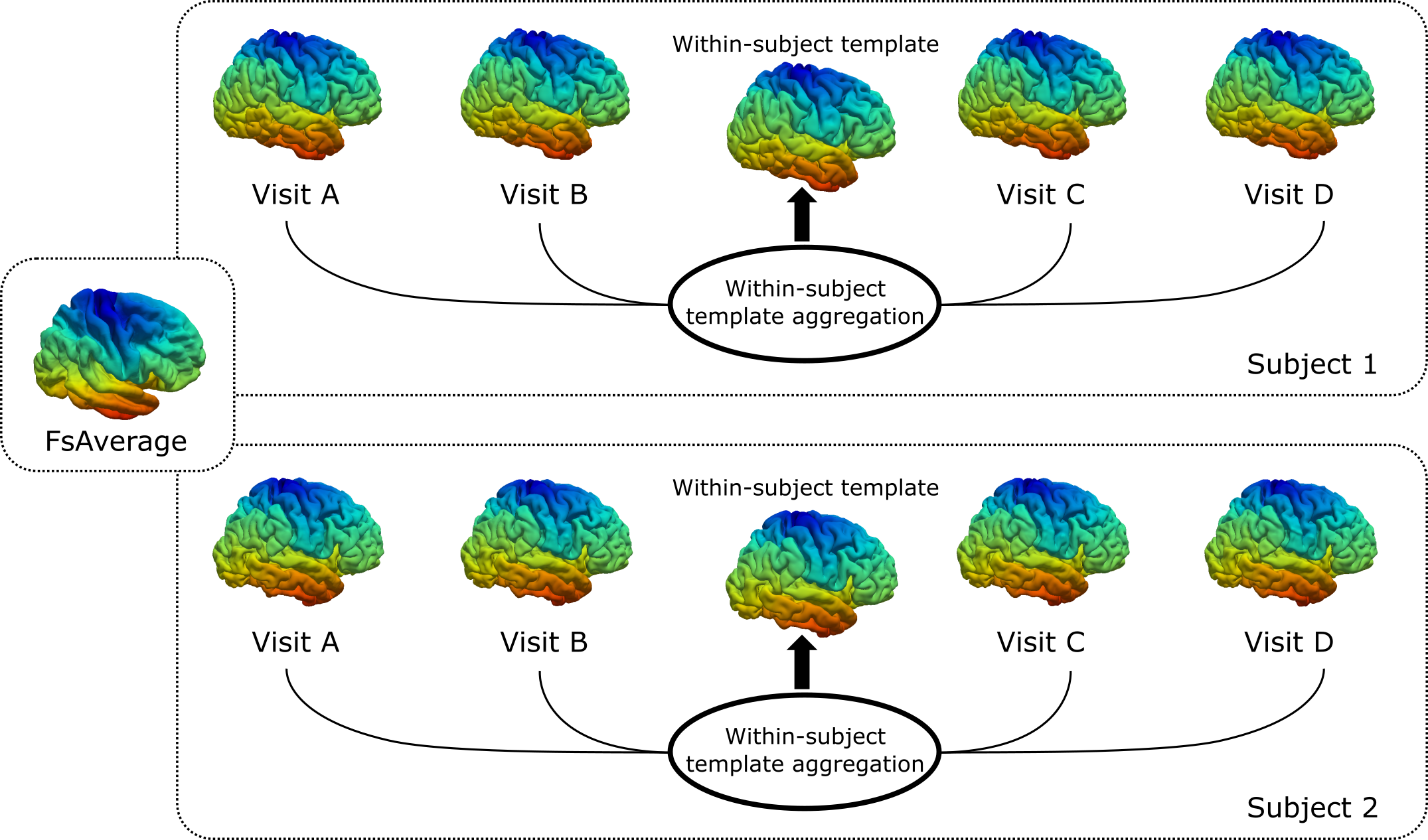}
    \caption{\rev Visualization of the correspondences to FsAverage that permit inter- and intra-subject averaging of vertices. We show pial surfaces, reconstructed with V2C-Flow, from two subjects in our ADNI test set and four visits each, together with the obtained within-subject template in V2C-Long. Every vertex is assigned an individual color so that correspondences can be identified by matching colors. }
    \label{fig:real-correspondences}
\end{figure}

\begin{figure}
    \centering
    \includegraphics[width=\textwidth]{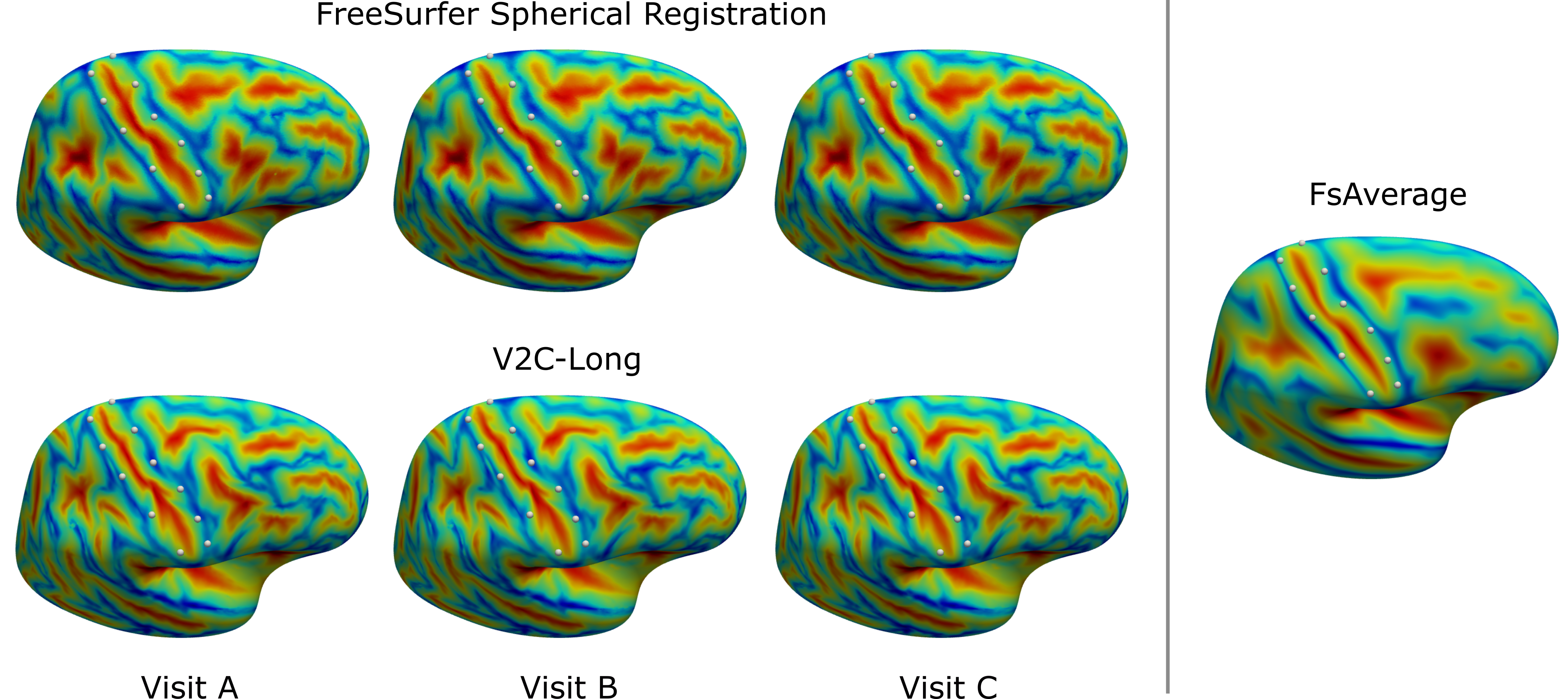}
    \caption{\revtwo Visualization of the within-subject correspondences and the correspondences to FsAverage based on sulcal depth maps. Shown is the inflated FsAverage mesh, with vertices around the central sulcus marked in white. After FreeSurfer's spherical registration, values were resampled to the FsAverage mesh; values from V2C-Long were mapped directly based on the inherent vertex correspondence. Ideally, sulcal depth values match at a certain vertex. The scans used for this analysis originate from our ADNI test set.}
    \label{fig:sulc-correspondences}
\end{figure}

\end{document}